\documentclass[%
 reprint,
superscriptaddress,
 amsmath,amssymb,
 aps,
pra,
]{revtex4-1}

\usepackage{titlesec}
\usepackage{amsthm,amssymb}
\usepackage{amsmath}
\usepackage{graphicx}
\usepackage{dcolumn}
\usepackage{bm}
\usepackage{multirow} 
\usepackage{mathdots}
\usepackage{enumerate}
\usepackage{algorithm}
\usepackage{algorithmicx}
\usepackage{algpseudocode}
\usepackage{textcomp}
\usepackage{color}

\usepackage{epstopdf}
\usepackage{epsfig}
\usepackage{array}
\usepackage{booktabs}
\usepackage{subfigure}%
\usepackage{setspace}%
\usepackage{amsmath}%
\usepackage{tikz}%
\usepackage{hyperref}
\usepackage{multirow}
\usepackage{changes}

\usetikzlibrary{shapes,arrows}

\theoremstyle{remark}

\begin{document}

\textbf{\textit{\subsubsection}}

\makeatletter
\newcommand{\ud}{\mathrm{d}}
\newcommand{\rmnum}[1]{\romannumeral #1}
\newcommand{\polylog}{\mathrm{polylog}}
\newcommand{\ket}[1]{|{#1}\rangle}
\newcommand{\bra}[1]{\langle{#1}|}
\newcommand{\inn}[2]{\langle{#1}|#2\rangle}
\newcommand{\Rmnum}[1]{\expandafter\@slowromancap\romannumeral #1@}
\newcommand{\udots}{\mathinner{\mskip1mu\raise1pt\vbox{\kern7pt\hbox{.}}
        \mskip2mu\raise4pt\hbox{.}\mskip2mu\raise7pt\hbox{.}\mskip1mu}}
\makeatother

\preprint{APS/123-QED}

\title{Progressive Quantum Algorithm for Maximum Independent Set with Quantum Alternating Operator Ansatz }

\author{Xiao-Hui Ni}
\affiliation{State Key Laboratory of Networking and Switching Technology, Beijing University of Posts and Telecommunications, Beijing, 100876, China}
\affiliation{School of Cyberspace Security, Beijing University of Posts and Telecommunications, Beijing, 100876, China}

\author{Ling-Xiao Li}
\affiliation{State Key Laboratory of Networking and Switching Technology, Beijing University of Posts and Telecommunications, Beijing, 100876, China}

\author{Yan-Qi Song}
\affiliation{China Academy of Information and Communications Technology, Beijing, China}

\author{Zheng-Ping Jin}
\affiliation{State Key Laboratory of Networking and Switching Technology, Beijing University of Posts and Telecommunications, Beijing, 100876, China}
\affiliation{School of Cyberspace Security, Beijing University of Posts and Telecommunications, Beijing, 100876, China}
\affiliation{National Engineering Research Center of Disaster Backup and Recovery, Beijing University of Posts and Telecommunications}

\author{Su-Juan Qin}
\email{qsujuan@bupt.edu.cn}
\affiliation{State Key Laboratory of Networking and Switching Technology, Beijing University of Posts and Telecommunications, Beijing, 100876, China}
\affiliation{School of Cyberspace Security, Beijing University of Posts and Telecommunications, Beijing, 100876, China}
\affiliation{National Engineering Research Center of Disaster Backup and Recovery, Beijing University of Posts and Telecommunications}

\author{Fei Gao}
\email{gaof@bupt.edu.cn}
\affiliation{State Key Laboratory of Networking and Switching Technology, Beijing University of Posts and Telecommunications, Beijing, 100876, China}
\affiliation{School of Cyberspace Security, Beijing University of Posts and Telecommunications, Beijing, 100876, China}
\affiliation{National Engineering Research Center of Disaster Backup and Recovery, Beijing University of Posts and Telecommunications}

\date{\today}

\begin{abstract}
Hadfield et al. proposed a novel Quantum Alternating Operator Ansatz algorithm (QAOA+), and this algorithm has wide applications in solving constrained combinatorial optimization problems (CCOPs) because of the advantages of QAOA+ ansatz in constructing a feasible solution space. In this paper, we propose a Progressive Quantum Algorithm (PQA) with QAOA+ ansatz to solve the Maximum Independent Set (MIS) problem using fewer qubits. The core idea of PQA is to construct a subgraph that is likely to contain the MIS solution of the target graph and then solve the MIS problem on this subgraph to obtain an approximate solution. To construct such a subgraph, PQA starts with a small-scale initial subgraph and progressively expands its graph size utilizing heuristic expansion strategies. After each expansion, PQA solves the MIS problem on the newly generated subgraph. In each run, PQA repeats the expansion and solving process until a predefined stopping condition is reached. Simulation results demonstrate that to achieve an approximation ratio of 0.95, PQA requires only $5.565\%$ ($2.170\%$) of the qubits and $17.59\%$ ($6.430\%$) of the runtime compared with directly solving the original problem using QAOA+ on Erdős-Rényi (3-regular) graphs, highlighting the efficiency of PQA.
\end{abstract}

\pacs{Valid PACS appear here}
\maketitle

\section{Introduction}\label{sec1}
Quantum computing has made significant strides in recent years, showing great potential in solving complex problems such as database searching \cite{grover}, prime factorization \cite{shor}, data dimension reduction \cite{qpca, reduction_yu, reduction_pan, reduction_wan}, cross-validation \cite{Lijing}, and solving equations \cite{hhl}. Despite these promising advancements, the practical implementation of quantum algorithms faces challenges due to the noise and limited qubit capacity of current Noisy Intermediate-Scale Quantum (NISQ) devices \cite{NISQ, VQA_review1, VQA_review2}. To fully harness the power of NISQ devices, hybrid quantum-classical algorithms \cite{adaptive-vqe, qubit-adaptive-vqe, QCL, QAS, RE-ADAPT-VQA, QFL, wu_quantum, wu_qst} have been proposed.

\medskip
The Quantum Approximate Optimization Algorithm (QAOA) \cite{qaoa, RQAOA, tree-QAOA, adaptive-qaoa} is a class of hybrid quantum-classical algorithms widely used to solve a variety of Combinatorial Optimization Problems (COPs) \cite{xiumei,exact_cover, INTERP, perform1, constraints_QAOA, portfolio, divide_MIS, MLI}, demonstrating its versatility in addressing both unconstrained and constrained optimization tasks. In the context of Constrained COPs (CCOPs), QAOA typically encodes problem constraints into the target Hamiltonian by introducing penalty terms \cite{minimum_vertex_cover, QIRO, Ising_formulations} and subsequently searches for the quasi-optimal solution using a Parameterized Quantum Circuit (PQC). However, this approach can result in infeasible solutions (i.e., solutions that violate one or more problem constraints) since QAOA explores the entire Hilbert space, including both feasible and infeasible states, without explicit constraints on the solution space \cite{RUAN}.

\medskip
To overcome these challenges, Hadfield et al. introduced a novel Quantum Alternating Operator Ansatz algorithm (QAOA+) \cite{alternating_QAOA_ansatz}, an enhanced version of the original QAOA. QAOA+ constructs a Hilbert space that includes only feasible solutions by encoding the problem constraints into the mixer Hamiltonian, and then searches for a quasi-optimal solution within this constrained space. This approach reduces the search space and eliminates the risk of infeasible solutions, offering a significant advantage over QAOA \cite{constraints_QAOA}. QAOA+ has been successfully applied to many CCOPs \cite{graph_coloring1, graph_coloring2, portfolio_optimization, MIS, minimum_exact_cover}, such as the Minimum Exact Cover problem. Although QAOA+ offers significant advantages in solving CCOPs, reducing the qubit requirements of QAOA+ is essential to tackle larger-scale CCOPs given the limited resources of current NISQ devices, thereby further expanding its applications.

\medskip
To achieve this goal, Tomesh et al. proposed a Quantum Local Search (QLS) algorithm with the QAOA+ ansatz and evaluated its performance on the Maximum Independent Set (MIS) problem \cite{QLS} that is a significant CCOP encountered in various fields, including network design \cite{MIS_in_network} and scheduling \cite{MIS_in_scheduling}, and is equivalent to the minimum vertex cover or finding a maximum clique in the complementary graph \cite{MIS_vertex_cover}. Unlike directly solving the MIS problem on the original graph using QAOA+ (DS-QAOA+) \cite{MIS}, QLS constructs a global solution by solving many local subproblems. Specifically, QLS constructs multiple local subgraphs using specific strategies and solves the MIS problem on each subgraph with QAOA+ to obtain local solutions. These local solutions are then combined into a global solution through classical methods. By breaking the original problem into multiple sub-problems, QLS reduces the number of qubits needed for solving the MIS problem on the original graph, making it more suitable for current quantum devices with limited resources. Although the qubit requirements decrease as the subgraph size reduces, the complexity of merging local solutions increases as the number of subgraphs grows.

\medskip
In this paper, we propose a quantum heuristic algorithm called the Progressive Quantum Algorithm (PQA), inspired by the Proactively Incremental Learning algorithm \cite{PIL} and the properties of the MIS problem, to reduce the qubit requirements of QAOA+ in solving the MIS problem. The core idea of PQA is to search for a subgraph that is likely to contain the MIS solution of the original graph and then solve the MIS problem on this subgraph to obtain an approximate solution. To construct such a subgraph, PQA commences with a small and sparse subgraph, gradually enlarging its size through strategies distinct from those utilized in QLS. Following each expansion, PQA employs QAOA+ to address the MIS problem on the freshly constructed subgraph. This process of graph expansion and problem resolution is iterated by PQA until a predetermined termination criterion is satisfied. By doing so, PQA can solve the original MIS problem in a smaller subgraph than DS-QAOA+. Additionally, unlike QLS, multiple local solutions are not required to be combined to reconstruct the global solution. Numerical results demonstrate that PQA generally achieves the same approximation ratio (AR) with fewer qubits and shorter runtime compared with DS-QAOA+ and QLS. Specifically, for an AR of 0.95, PQA uses $5.57\%$ of the qubits required by DS-QAOA+ and $26.77\%$ of those required by QLS on ER graphs, and $2.17\%$ and $21.76\%$ on 3-regular graphs, respectively. Additionally, PQA achieves the same AR in $17.59\%$ of the runtime required by DS-QAOA+ and $76.72\%$ of that required by QLS on ER graphs, and $6.43\%$ and $61.49\%$ on 3-regular graphs. The efficiency of PQA in solving the MIS problem can enhance the practicality of quantum algorithms and open up possibilities for solving large-scale problems in the future.

\medskip
This paper is structured as follows. Section~\ref{sec:rev} serves as the foundational segment, providing background knowledge on the MIS problem, QAOA, and its extension, QAOA+. Following this, in Section~\ref{sec:PQA}, we delineate the progressive quantum algorithm, emphasizing the construction strategies of the desired derived subgraph. Subsequently, in Section~\ref{simulation}, we present a comprehensive set of numerical simulation experiments aimed at showcasing the quantum resource efficiency of PQA over DS-QAOA+ and QLS. Finally, in Section~\ref{conclusion}, we offer a succinct conclusion and discussion to encapsulate the key findings of our study and the next research direction.

\section{Preliminaries}\label{sec:rev}
To help readers understand our work better, we first introduce the MIS problem in Section~\ref{MIS}, followed by an introduction to QAOA in Section~\ref{QAOA}. Finally, we review the QAOA+ algorithm in Section~\ref{QAOA+}.

\subsection{Review of MIS}\label{MIS}
The MIS is defined on a graph $G=(V, E)$, where $V = \left \{ {0,\dots,n-1} \right \}$ is the set of vertices and $E = \{ \{ u, v \} \mid u, v \in V, u \neq v \}$ is the set of edges. In the graph theory, two vertices $u$ and $v$ are considered adjacent if they are connected by an edge $(u,v)$, and a subset of vertices $V_s$ is called an independent set of $G$ if no two vertices in $V_s$ are adjacent. The MIS is the vertex set with the most number of vertices among all independent sets \cite{reducibility}, and the number of vertices in the MIS is known as the independence number, denoted as $\beta(G)$. While the solution to the MIS problem for a graph may not be unique, the independence number is always uniform.

\begin{figure}[htbp]
	\centering
	\setlength{\abovecaptionskip}{0.cm}
	\subfigure[]{
    \includegraphics[width=0.43\linewidth]{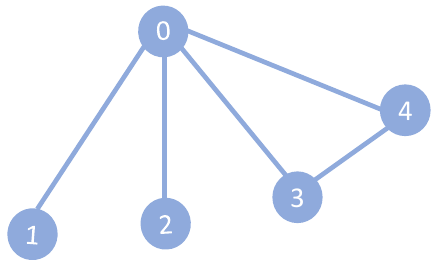}
    }
    \hspace{0.05cm} 
    \subfigure[]{
    \includegraphics[width=0.43\linewidth]{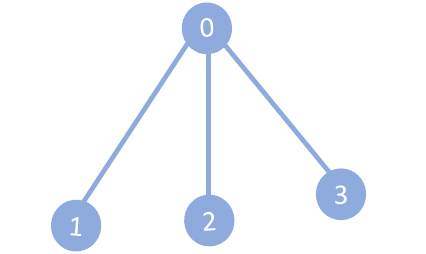}
    }
   \caption{Examples of MIS. (a) A target graph $G$ whose feasible solutions are $\varnothing$, $\left \{0\right \}$, $\left \{1\right \}$, $\left \{2 \right \}$, $\left \{3  \right \}$, $\left \{4  \right \}$, $\left \{1,2 \right \}$, $\left \{1,3  \right \}$, $\left \{1,4  \right \}$, $\left \{2,3  \right \}$, $\left \{2,4 \right \}$, $\left \{1,2,3  \right \}$ and $\left \{1,2,4  \right \}$. The solutions of MIS are $\left \{1,2,3  \right \}$ and $\left \{ 1,2,4 \right \}$, where the MIS solutions of $G$ are not unique, but the independence number identically equals $3$. (b) An induced subgraph of $G$ and its MIS solution is $\left \{ 1,2,3 \right \}$. The feasible solutions of $G_{q}$ are $\varnothing$, $\left \{0\right \}$, $\left \{1\right \}$, $\left \{2 \right \}$, $\left \{3  \right \}$, $\left \{1,2\right \}$, $\left \{1,3\right \}$, $\left \{2,3 \right \}$, $\left \{1,2,3  \right \}$. We respectively denote the feasible space including all feasible solutions of $G_{q}$ and $G$ as $S(G_{q})$ and $S(G)$, where $S(G_{q}) \subseteq S(G)$.}\label{MIS_example}
\end{figure}

\begin{figure*}[t]
	\centering
	\setlength{\abovecaptionskip}{0.cm}
	\subfigure{
		\includegraphics[width=0.99\textwidth]{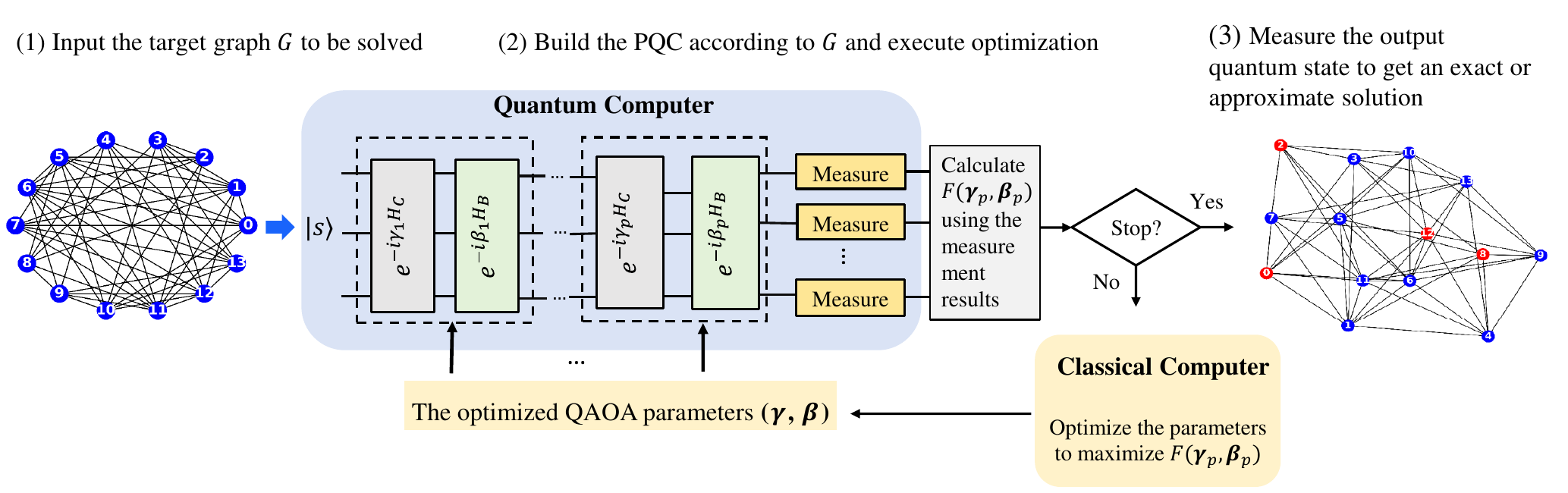}}
    \caption{A schematic of solving the MIS problem on the target graph using QAOA.}
\label{direct_solving}
\end{figure*}

\medskip
For any target graph $G$, the independent vertex subsets in an induced subgraph $G_{q}$ of $G$ are feasible solutions for $G$. That is, there is at least one induced subgraph $G_{q} = (V_{q},E_{q})$ that has an identical MIS solution to $G$, where $V_{q} \subseteq V$ and $E_{q} \subseteq E$. As illustrated in FIG.~\ref{MIS_example}, the right graph is the induced subgraph of the left graph, and they have the same MIS solution $\left \{ 1,2,3 \right \}$, but the MIS solution $\left \{ 1,2,4 \right \}$ to the target graph is not the MIS solution of the right graph because the node $4$ is not in this induced subgraph, which implies that the MIS solution of the target graph may not always be a solution for its induced subgraph.

\medskip
In this paper, we denote the number of vertices and edges in $G$ as $n$ and $m$, respectively. To formalize the MIS problem, a binary variable $x_{u}$ for each vertex $u$ in $V$ is introduced, where $x_{u} = 1$ represents that the vertex $u$ is in the vertex subset, otherwise, $x_{u} = 0$. Each way of vertex partition corresponds to a unique bit string $x$, and there are $2^{n}$ forms for $x$ in total, where $x = x_{0}x_{1} \dots x_{n-1}$. The goal of the MIS problem is to maximize the following objective function
\begin{equation}
	C_{1}(x) = \sum_{u = 0}^{n-1}x_{u}\label{goal}
\end{equation} 
when the independence constraint $\sum_{(u,v) \in E} x_{u}x_{v} = 0$ is met. This constraint can be incorporated into the objective function using the Lagrange multiplier method \cite{Iterative_MIS}, thus converting the constrained problem to an unconstrained problem. The obtained classical cost function is
\begin{equation}
	\max  C(x) = \sum_{u = 0}^{n-1}x_{u} - \lambda \sum_{(u,v) \in E} x_{u}x_{v},\label{cost_ function}
\end{equation} 
where the value of $\lambda$ is determined through additional hyper-parameter optimization steps \cite{QIRO}, and $\lambda > 1$ can impose penalties on those terms that violate independence constraints \cite{Ising_formulations}, making the maximal cost function value correspond to the exact MIS solution instead of other feasible or infeasible bit strings. Here, $C_{\text{max}}$ equals the independence number $\beta (G)$. 

\medskip
The MIS problem has been rigorously proven to be NP-hard. As the number of vertices in a graph increases, the time complexity of solving the MIS problem using exact algorithms, such as backtracking \cite{backtracking_algorithm_MIS} and branch-and-bound methods \cite{branch_and_bound_MIS}, grows exponentially \cite{exact_algorithm_MIS}. This rapid escalation in computational demands renders exact algorithms increasingly impractical for large-scale instances. To address this challenge, researchers have proposed heuristic algorithms. Some of the representative classical heuristic approaches include evolutionary algorithms \cite{evolutionary_algorithm_MIS}, tabu search \cite{tabu_search}, and local search \cite{hybrid_iterated_LS,iterated_LS,local_search_MIS}. Although these algorithms sacrifice some degree of precision compared with the exact algorithms, they may provide quasi-optimal solutions in a shorter time, striking a balance between efficiency and accuracy.

\subsection{Review of QAOA}\label{QAOA}
QAOA is an approximate algorithm and provides an approximation solution to the solved problem. QAOA encodes the solution of the problem into the ground state of the target Hamiltonian $H_{C}$ and aims to approximate this ground state through a PQC \cite{qaoa}. This evolution starts from the ground state $|s\rangle$ of the initial Hamiltonian $H_{B}$, with a quantum circuit built using $p$-level QAOA ansatz, where $p$ is the level depth. Notably, the initial state $|s\rangle$ is required to be trivial to implement. That is, $|s\rangle$ can be prepared by a constant-depth quantum circuit. 

\medskip
For the original QAOA, one level QAOA ansatz consists of two variational QAOA parameters (denoted as $\gamma_{i}$ and $\beta_{i}$) and two unitaries $e^{-i\gamma_{i}H_{C}}$ and $e^{-i\beta_{i}H_{B}}$, where $i = 1,2,\cdots,p$. Generally, the initial Hamiltonian is chosen as $H_{B}= -\sum_{u=0}^{n-1} \sigma _{u}^{x}$ because its ground state $|s\rangle = |+\rangle ^{\otimes n}$ can be trivial to prepare. Here, $\sigma_{u}^{x}$ refers to the Pauli-X operator applied to the 
$u-$th qubit, effectively flipping its state. To encode the MIS solution into the ground state of $H_{C}$, the cost function $-C(x)$ is transformed into the target Hamiltonian
\begin{equation}
	H_{C}=\sum_{u = 0}^{n-1}\frac{\sigma_{u}^{z}-I}{2} + \lambda \sum_{(u,v) \in E}\frac{I-\sigma_{u}^{z}-\sigma_{v}^{z}+\sigma_{u}^{z}\sigma_{v}^{z} }{4}\label{target_ham}
\end{equation} 
by transforming each binary variable $x_{u}$ to a quantum spin $\frac{I-\sigma_{u}^{z} }{2} $ \cite{qaoa}, where $\sigma_{u}^{z}$ refers that applying Pauli-Z to the $u$-th qubit.

\medskip
The ground state of $H_{C}$ can be approximately obtained by alternately applying unitaries operators $e^{-i\gamma _{i} H_{C}}$ and $e^{-i\beta _{i}  H_{B} }$ to the initial quantum state $|s\rangle$. After applying these unitaries for 
$p$ levels, the output state of the PQC is 
\begin{equation}
	|\boldsymbol{\gamma_{p}},\boldsymbol{\beta_{p}}\rangle  =\prod_{i=1}^{p} e^{-i\beta _{i}  H_{B} }e^{-i\gamma _{i}  H_{C} }, \label{quantum_state}
\end{equation} 
where $\boldsymbol{\gamma_{p}} = (\gamma_{1},...,\gamma_{p})$ and $\boldsymbol{\beta_{p}} = (\beta_{1},...,\beta_{p})$ are $2p$ variational QAOA parameters. The expectation function value of $H_{C}$ in this variational quantum state is defined as
\begin{equation}
	F({\boldsymbol{\gamma_{p}}}  ,{\boldsymbol{\beta_{p}}})	= - \langle {\boldsymbol{\gamma_{p}}},{\boldsymbol{\beta_{p}}} |H_{C}|{\boldsymbol{\gamma_{p}}}  ,{\boldsymbol{\beta_{p}}}\rangle ,\label{expectation}
\end{equation} 
which can be calculated by repeated measurements of the output quantum state using the computational basis. To search for the global optimal parameters 
\begin{equation}
	(\boldsymbol{\gamma_{p}}^{\text{opt}},\boldsymbol{\beta_{p}}^{\text{opt}}) = \arg  \underset{(\boldsymbol{\gamma_{p}},\boldsymbol{\beta_{p}})}{\max} F(\boldsymbol{\gamma_{p}},\boldsymbol{\beta_{p}}),
\end{equation} 
multiple optimization runs tend to be required. In each run, the search is typically done by starting with some guesses of the initial parameters. Then, the quantum computer measures the output state to compute the expected value of the output state and delivers the relevant information to the classical computer. These parameters are optimized by the classical optimizer and then delivered to the quantum computer. The outer parameter optimization is repeated until it meets the termination condition, such as the maximal number of iterations or convergence tolerance $\varepsilon$ \cite{MLI}.  

\medskip
To quantify the quality of the QAOA solution (i.e., measure the difference between the output state of QAOA and the ground state of $H_{C}$), the approximation ratio (AR)
\begin{equation}
	r = \frac{F({\boldsymbol{\gamma_{p}}}  ,{\boldsymbol{\beta_{p}}})}{F_{\text{max}} } \label{ratio}
\end{equation} 
is introduced, where $F_{\text{max}}$ is the negative of the ground state energy of $H_{C}$, with equaling $C_{\text{max}}$. For small graph instances, \( C_{\text{max}} \) can be obtained by brute force search. For large graphs, an approximation of \( C_{\text{max}} \) can be obtained by executing multiple runs of the classical heuristic algorithm and selecting the largest independent set found across these runs. AR reflects how close the solution provided by QAOA is to the exact solution, $r \le 1$, with the value of 1 as the exact solution.  Finally, to help readers better understand the QAOA, its detailed process is shown in FIG.~\ref{direct_solving}.

\begin{table*}
	\caption  { \centering The difference between QAOA and QAOA+ when solving the MIS problem}
	\label{difference}%
	\begin{ruledtabular}
		\renewcommand\arraystretch{1.5}
		\begin{tabular*}{\textwidth}{@{}ccc@{}}
			\toprule
			\textbf{} &\textbf{QAOA} &\textbf{QAOA+}   \\ 
			\hline
			\text{Manipulation of constraint}  &\text{Add it into $H_{C}$ by Lagrange multiplier} &\text{Encode it into $H_{B}$}  \\
			\text{Initial state $|s\rangle$} 		&\text{$|s\rangle  = |+\rangle ^{\otimes n} $}  &\text{ The superposition of feasible solutions} \\
			\text{Mixer Hamiltonian $H_{B}$} 		&\text{$H_{B}= -\sum_{u=0}^{n-1} \sigma _{u}^{x}$}  &\text{$H_{B} = \sum_{u = 0}^{n-1}B_{u}$} \\
			\text{Target Hamiltonian $H_{C}$} 		&\text{$ H_{C} = \sum_{u = 0}^{n-1}\frac{\sigma_{u}^{z}-I}{2} + \lambda \sum_{(u,v) \in E}\frac{I-\sigma_{u}^{z}-\sigma_{v}^{z}+\sigma_{u}^{z}\sigma_{v}^{z} }{4}$}  &\text{$H_{C} = \sum_{u = 0}^{n-1}\frac{\sigma_{u}^{z}-I}{2}$} \\
			\text{$|s\rangle$ is the ground state of $H_{B}$ ?} &\text{Yes} &\text{No requirement} \\
			\text{Hilbert space} &\text{Including feasible and infeasible states} &\text{Feasible states}\\
			\bottomrule
		\end{tabular*}
	\end{ruledtabular}
\end{table*}

\subsection{Review of QAOA+}\label{QAOA+}
The original QAOA searches for the ground state of $H_{C}$ from a Hilbert space that includes both feasible and infeasible quantum states, the output states may not always satisfy the independence constraint and violate one or more constraints \cite{constraints_QAOA, QIRO}. To overcome this challenge, Hadfield et al.~\cite{graph_coloring1,alternating_QAOA_ansatz} introduce a novel quantum alternating operator ansatz algorithm denoted as QAOA+.

\medskip
Different from incorporating the independence constraint into the target Hamiltonian utilizing the Lagrange multiplier, QAOA+ encodes constraints into the mixer Hamiltonian $H_{B}$. This allows QAOA+ to search for the optimal solution within a subspace that only contains feasible quantum states. Consequently, the target Hamiltonian 
\begin{equation}
	H_{C} = \sum_{u = 0}^{n-1}\frac{\sigma_{u}^{z}-I}{2}\label{H_C,QAOA+}
\end{equation}
is obtained by converting the $x_{u}$ in $-C_{1}(x)$ to a quantum spin $\frac{I-\sigma_{u}^{z} }{2} $, thereby encoding the MIS solution into the ground state of $H_{C}$. Similar to QAOA, QAOA+ approaches the ground state of $H_{C}$ by alternately applying target and mixer unitaries on an initial quantum state $|s\rangle$ that is required to be trivial to implement. However, in QAOA+, the initial quantum state must be the superposition of feasible states. This requirement ensures that the feasible subspace contains no states that violate the constraints. In addition, the initial quantum state $|s\rangle$ may not be the ground state of $H_{B}$ in QAOA+, unlike in QAOA.

\begin{figure}
	\centering
	\setlength{\abovecaptionskip}{0.cm}
	\subfigure{
		\includegraphics[width=0.48\textwidth]{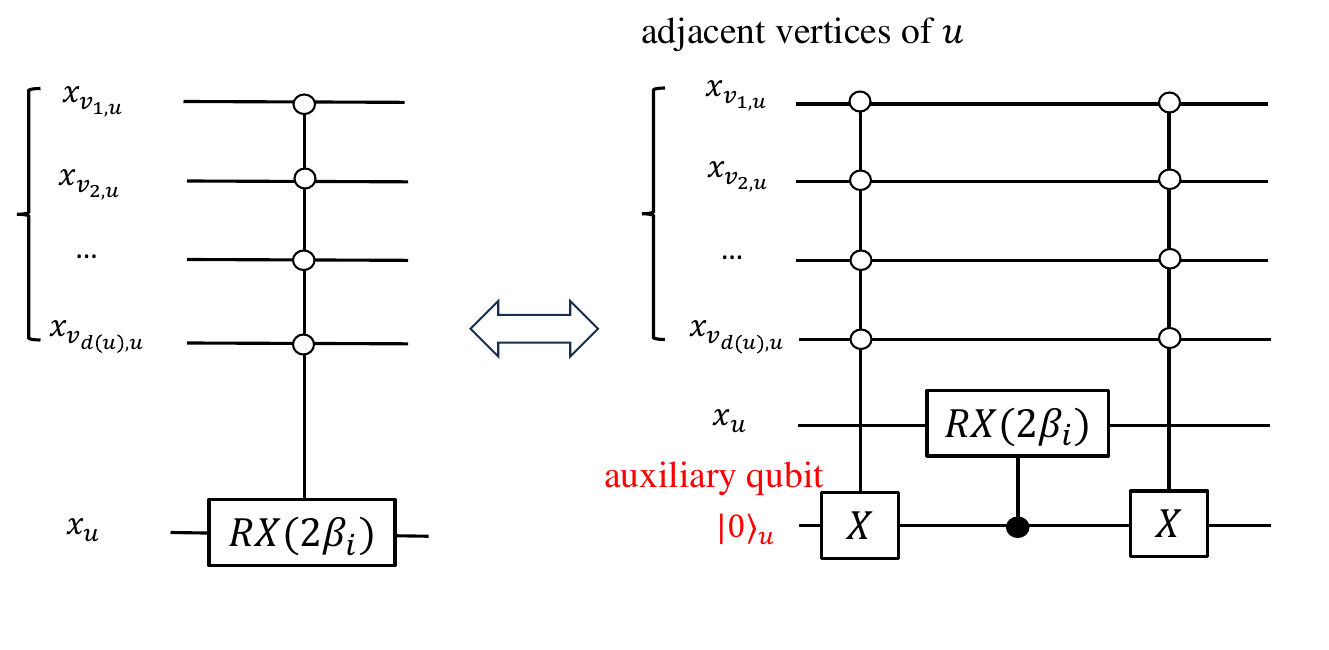}
	}\caption{The implementation of $e^{-i\beta_{i}B_{u}}$ when $d(u) \ge 1$. The multi-qubit controlled $R_{x}$ gate can be decomposed into two multi-qubit controlled gates and a single-qubit controlled $R_{x}$ gate, utilizing an auxiliary qubit \cite{gate_decomposition}. The auxiliary qubit is initialized as $|0\rangle$ and is flipped to $|1\rangle$ only when the quantum states of all adjacent vertices are $|0\rangle$ (i.e., $f(u) = 0$). Once this condition is met, an $R_{x}$ gate is applied on the $u$-th qubit. After the operation, the state of the auxiliary qubit can be reset from $|1\rangle$ to $|0\rangle$ using the right multi-qubit controlled gate operation. Otherwise, the auxiliary qubit is always $|0\rangle$ when $f(u) >0$, and there is an Identity operation applied to the $u$-th qubit.}\label{H_B_circuit}
\end{figure}

\medskip
The design of the mixing Hamiltonian $H_{B}$ must preserve the feasibility of the subspace (i.e., ensure the subspace only has quantum states obeying constraints) and explore a new feasible subspace (i.e., provide transitions between various feasible states) \cite{alternating_QAOA_ansatz}. For an independent vertex subset $V_{s}$, deleting an arbitrary vertice $v_{a}$ (i.e., the state of $x_{v_{a}}$ is converted from 1 to 0) from $V_{s}$ does not change its independence. However, a new vertex $v_{a}$ can be added to $V_{s}$ (i.e., flipping $x_{v_{a}}$ from 0 to 1) without destroying its independence only when there are no adjacent vertices of $v_{a}$ in $V_{s}$. Thus, for any node, controlling its bit-flip operation based on the states of its neighbors is sufficient to both add and remove vertices while preserving the independence property \cite{alternating_QAOA_ansatz}. The bit-flip of $v_{a}$ occurs only if 
\begin{equation}
	f(v_{a}) = \sum_{j = 1}^{d(v_{a})}x_{v_{j,a}} = 0,
\end{equation}
where $v_{j,a}$ is the $j$-th adjacent vertex of $v_{a}$, and $d(v_{a})$ is the number of adjacent vertices of $v_{a}$, with $j = 1,\cdots,d(v_{a})$. Mapping to a quantum circuit, for any vertex $u$, a Pauli-X operation is applied on the $u$-th qubit (i.e., this qubit is flipped) if and only if all quantum states of adjacent vertices of $u$ are $|0\rangle$ (i.e., $f(u) = 0$). Otherwise, a Pauli-I operation is applied on the $u$-th qubit. For the node $u$, its corresponding partial mixer Hamiltonian $B_{u}$ can be described as
\begin{equation}
	\begin{split}
		B_{u} = \sum_{f(u) = 0} |x_{v_{1,u}} \cdots x_{v_{d(u),u}}\rangle \langle x_{v_{1,u}}\cdots x_{v_{d(u),u}}| \otimes  \sigma_{u}^{x}
		\\
		+ \sum_{f(u) > 0} |x_{v_{1,u}}\cdots x_{v_{d(u),u}}\rangle \langle x_{v_{1,u}}\cdots x_{v_{d(u),u}}| \otimes I.
	\end{split}
\end{equation}
The mixer Hamiltonian is given by $H_{B} = \sum_{u = 0}^{n-1}B_{u}$ \cite{alternating_QAOA_ansatz}. Specifically, the quantum state resulting from applying $e^{-i\beta_{i}B_{u}}$ to the single feasible quantum state $|x_{o}\rangle$ is
\begin{equation}
		e^{-i\beta_{i}B_{u}}|x_{o}\rangle = \begin{cases}
			&R_{x}(2\beta_{i})|x_{o}\rangle, f(u) = 0\\
			&|x_{o}\rangle, f(u)>0,
		\end{cases} \label{action_effect}
\end{equation}
where the detailed result of $R_{x}(2\beta_{i})|x_{o}\rangle$ is $\cos 2\beta_{i}|x_{o}\rangle - i\sin 2\beta_{i}|x_{0}\cdots \bar{x_{u}} \cdots x_{n-1}\rangle$, and $\bar{x_{u}}$ denotes the state of $u$-th qubit is converted from $|0\rangle$ to $|1\rangle$ (or ($|1\rangle$ to $|0\rangle$), implementing the addition or removal of node $u$ without generating infeasible quantum states. The circuit implementation of $e^{-i\beta_{i}B_{u}}$ is shown in FIG.~\ref{H_B_circuit} when the node has $d(u) \ge 1$ adjacent vertices. Additionally, Table~\ref{difference} is provided to illustrate clearly the differences between using QAOA and QAOA+ to solve the MIS problem.

\section{Progressive quantum algotirhm}\label{sec:PQA}

Any arbitrary graph has at least one induced subgraph that preserves the same MIS solution as the original graph. Inspired by this property, PQA aims to iteratively construct a derived subgraph that is likely to contain the MIS solution of $G$ while being smaller in size. Naturally, solving the MIS problem on this subgraph requires fewer qubits than directly solving it on the original graph. To construct such an induced subgraph, PQA iteratively expands the subgraph using heuristic expansion strategies designed for this purpose, starting from a sparse and small subgraph. Once the subgraph has been expanded, PQA proceeds to tackle the MIS problem within this subgraph by utilizing QAOA+. Subsequently, PQA iterates the processes of graph expansion and problem-solving until a previously specified termination condition is fulfilled. In PQA, the purpose of the graph expansion is to improve the probability that the constructed subgraph contains the MIS solution of the original target graph, while solving the MIS problem on each subgraph using QAOA+ serves to estimate whether the current subgraph contains the MIS solution of the original graph by obtaining the expected function values. The expansion strategies and the setting of the stopping condition will be detailed in the following.

\subsection{The graph construction and expansion strategies}\label{construction_rules}

The original graph may have multiple derived subgraphs, each differing in size but retaining the same MIS solution. The overhead of implementing the multi-qubit controlled $R_{x}$ gate is expensive and increases polynomially as the number of control qubits (that correspond to the neighborhood nodes) grows \cite{gate_decomposition,gate_decomposition_MIS}. To minimize the required qubits and reduce the overhead of implementing multi-qubit controlled $R_{x}$ gates, \textbf{an ideal subgraph should not only have the same MIS solution as the target graph but also be small and sparse. Here, the small property helps reduce the number of qubits required, and the sparse property reduces the overheads associated with multi-qubit controlled gates.}

\medskip
To construct such an ideal subgraph, PQA starts from a sparse and small-sized initial subgraph $G_{0}$ and gradually expands it. During each expansion, the process favors adding the node with the lowest ``closeness'' to minimize the effect of the new node on the sparsity of the subgraph. The closeness is measured by a classical objective function
\begin{equation}
	C(v_{c}) = \sum_{v_{c} \in V_{c}, v_{q} \in V_{q}} E(v_{c}, v_{q}), \label{cost_function}
\end{equation}
where $V_{c} = V - V_{q}$ is the vertex candidate set, and $V_{q}$ is the vertex set of the current induced subgraph $G_{q}$. If the nodes $v_{c}$ and $v_{q}$ are adjacent (i.e., $(v_{c},v_{q}) \in E$), $E(v_{c}, v_{q}) = 1$. Otherwise, $E(v_{c}, v_{q}) = 0$. The value of $C(v_{c})$ measures how the addition of vertex $v_{c}$ to the current subgraph $G_{q}$ affects its sparsity, and a small value of $C(v_{c})$ indicates a minimal change in sparsity. 

\medskip

\begin{figure*}[htbp]
	\centering
	\setlength{\abovecaptionskip}{0.cm}
	\subfigure{
		\includegraphics[width=0.98\textwidth]{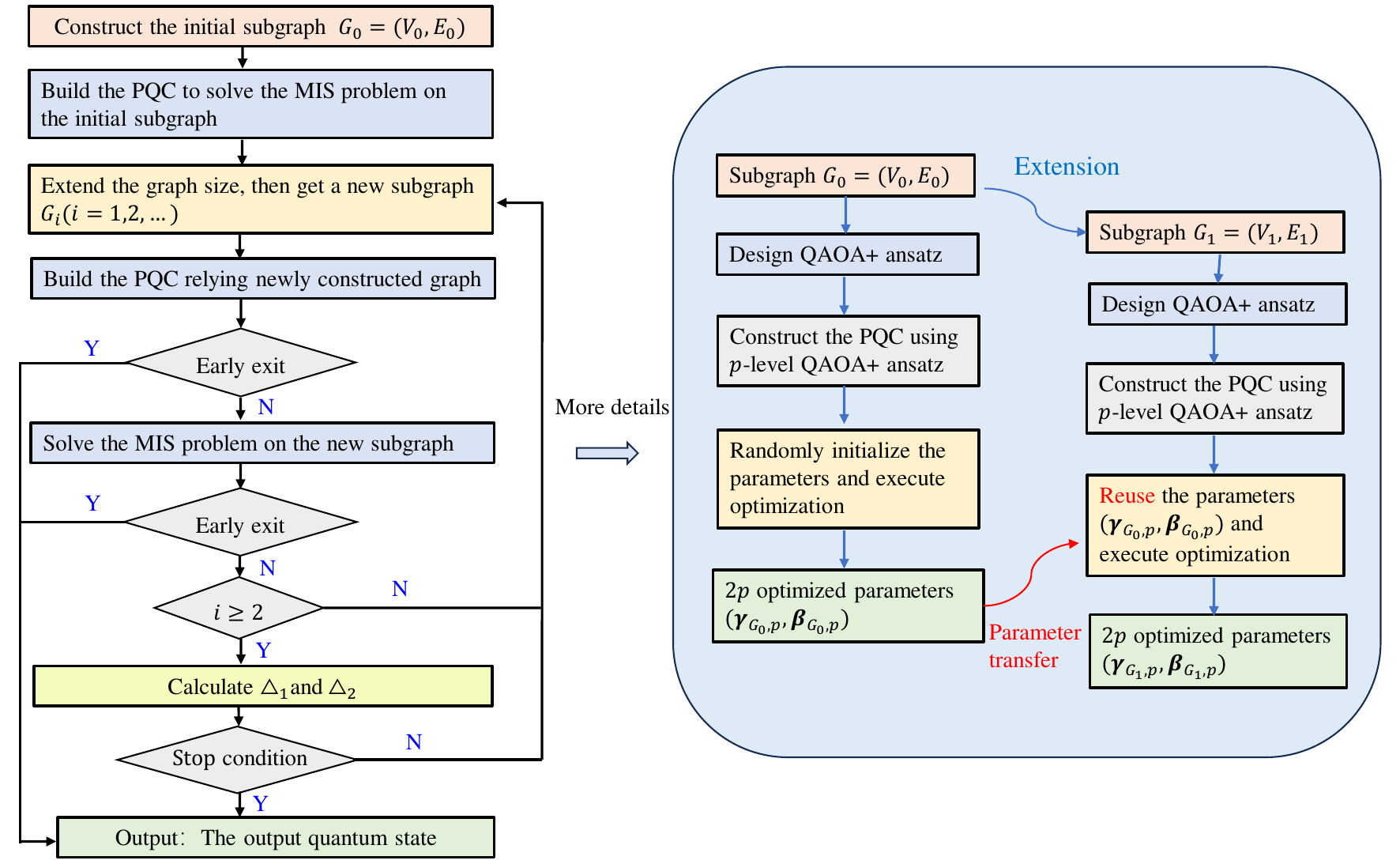}}
	\caption{A procedure schematic for PQA. PQA starts with an initial subgraph $G_{0}$ and solves the MIS problem using $p$-level QAOA+ ansatz, where the parameters of PQC are randomly initialized. Then, PQA progressively enlarges the graph and solves the MIS problem on the updated subgraph, retraining the PQC with pre-trained parameters. Notably, PQA incorporates two early exit mechanisms to terminate ineffective optimization, conserving quantum resources. The first mechanism assesses whether the current optimization process should proceed. The second mechanism evaluates whether further graph expansion is necessary. If the first mechanism is triggered, the PQC outputs the quantum state obtained from the previous subgraph $G_{i-1}$. The output state that corresponds to the maximal $F_{G_{j},p}$ ($j = 0, 1,...,i$) is given when the second mechanism is triggered. Otherwise, the quantum state is obtained from the current subgraph $G_i$ when the stop condition is met.}	\label{PQA}
\end{figure*}

\medskip
By starting with a small subgraph and gradually expanding its size, PQA can effectively control the growth of qubits, enhance computational efficiency, and minimize unnecessary resource usage. The construction of $G_{0}$ is not random, and it also involves the selection of the initial node and graph expansion (i.e., the addition of new vertices and edges). The detailed strategies are as follows. Additionally, an example is provided in Appendix~\ref{graph_extension} to help readers understand the process of graph expansion. 

\medskip
\textbf{The choice of the first node.} To construct a sparse initial subgraph $G_{0}$, the node with a minimum degree in the target graph $G$ is chosen as the first node of $G_{0}$. If there are $l (>1)$ nodes with a minimum degree, one is randomly chosen as the first node.

\medskip
\textbf{The extension of subgraph.} The addition of new nodes should have minimal impact on graph sparsity. That is, the node that has  $\min C(v_{c})$ in the candidate set is the next vertex to join the current subgraph. If there are $l (>1)$ nodes with $\min C(v_{c})$, further calculations are performed. Specifically, for each node with $\min C(v_{c})$, we assume it is added to the current subgraph and calculate its corresponding $\min C(v_{c})$, resulting in $l$ calculation results. The node with the minimum value among these results is chosen to expand the current subgraph. If multiple nodes meet this criteria, one is randomly selected as the next node. During the extension, new edges may be introduced. Specifically, any edge $(v_{q},v_{new})$ should be added to the current subgraph if the newly added node $v_{new}$ is adjacent to existing nodes $v_{q}$, resulting in an expanded subgraph.

\subsection{Progressive quantum optimization}
After constructing an initial induced subgraph $G_{0}$, PQA builds a PQC using the $p$-level QAOA+ ansatz to solve the MIS problem on $G_{0}$, with the parameters randomly initialized. After optimization, PQA continues to expand the graph size according to the given expansion strategies and constructs $p$-level QAOA+ ansatz for the new subgraph to solve the MIS problem. At this stage, the PQC reuses the optimized QAOA parameters from the previous subgraph and re-optimizes these pre-trained parameters. This process is referred to as ``parameter transfer'' \cite{PIL,parameter_transferability,parameter_transfer,transfer_learning2024,angle_conjecture, leapfrogging}. After each optimization round, PQA gets an expectation function value $F_{G_{i},p}$ about the subgraph $G_{i}$, and PQA terminates if the values satisfy $\Delta_{1} \le \xi$ and $\Delta_{2} \le \xi$ or if the size matches that of the target graph in the worst case, where
\begin{equation}
	\begin{aligned}
		\Delta_{1} = \lvert  F_{G_{i+1},p} - F_{G_{i},p} \rvert,\\
		\Delta_{2} = \lvert F_{G_{i},p} - F_{G_{i-1},p} \rvert.
	\end{aligned} \label{stop_condition}
\end{equation}
Here, $\xi$ is the tolerance for function values, and $F_{G_{i},p}$ ($F_{G_{i-1},p}$) represents the function value obtained using $p$-level QAOA+ ansatz to solve the MIS problem on the current (previous) subgraph $G_{i}$ ($G_{i-1}$).

\medskip
Here, we explain why $\Delta_{1} \le \xi$ and $\Delta_{2} \le \xi$ are used as the termination conditions. According to the properties of the MIS problem, adding a new vertex to the derived subgraph $G_{i}$ can cause the newly constructed subgraph $G_{i+1}$ to have an identical or larger independence number compared with $G_{i}$. In other words, $\beta(G_{i}) \le \beta(G_{i+1})$ (i.e., $F_{\text{max}, G_{i}} \le F_{\text{max}, G_{i+1}}$). When PQA starts with a subgraph that has the same MIS solution as the target graph, the growth in the maximal expected function values $F_{\text{max}, G_{i}}$ will cease as the subgraph grows (i.e., $F_{\text{max}, G_{i}}$ reaches a steady value). Nevertheless, the value of $F_{\text{max}, G_{i}}$ might also ascend with the increase of the graph size after a brief period of stagnation when PQA starts with a different initial subgraph. During the transient stationary phase, the addition of new nodes provides no benefit (i.e., without incrementing the maximal expectation function value). This occurs due to being trapped in a local subgraph, which is affected by sub-optimal node selection. For the latter case, the subsequent enlargement of the graph might enhance performance, but it concurrently requires more quantum resources to construct the corresponding quantum circuits. At this time, scaling the graph size without regard to resources is not advisable, and PQA should terminate the process in a timely manner. Given these considerations, we prioritize the stable phase of change in the expected function value during optimization, and PQA is halted if the expected function value remains stable over three consecutive optimization rounds. By this stage, we may have discovered a derived subgraph that includes the MIS solution of the target graph, obtaining an approximate or exact solution by solving the MIS problem on this subgraph.

\medskip
During actual optimization, we find the expectation function value may oscillate back and forth with increasing subgraph size. In such cases, PQA may not satisfy the termination condition and continue to execute optimization until reaching a similar scale to the target graph, even though a subgraph with the same solution as the target graph has been constructed. This contradicts the original goal of PQA. \textbf{To avoid unnecessary waste of resources in the optimization run, we set up early exit mechanisms that are required to end the low-cost optimization when appropriate. These mechanisms are based on evaluating the quality of the reused parameters and the optimization results for each subgraph.} After each graph extension, we first decide whether to proceed with the optimization round by assessing the quality of the reused (initial) parameters. Generally, PQA can converge to quasi-optima after a few iterations (or even no optimization) when it starts with high-quality initial parameters. Otherwise, it may require abundant iterations to converge or get stuck in a low-quality local solution. Thus, we terminate PQA early and skip the current optimization round if the expected function value, based on the initial parameters, is significantly lower than that obtained from the previous optimization. If the quality of the initial parameters meets the condition we set, a new round of early exit judgments is carried out after the optimization. If the latest expected function value is significantly lower than the maximal value obtained from previous subgraphs, it indicates that the optimization has fallen into a poor local solution. At this time, PQA will halt further graph expansion and optimization, returning the maximal function value in this run and the corresponding output state. Otherwise, PQA continues to execute if it satisfies the above settings. \textbf{In summary, to reduce the optimization resources required during a PQA run, we introduced two early exit mechanisms that determine whether subsequent optimization rounds are necessary based on `flags' (the quality of the initial parameters and the optimization results).} More detailed procedures of PQA are shown in FIG.~\ref{PQA}, which may help readers more intuitively understand the PQA process.

\medskip
PQA is a hybrid quantum-classical algorithm that integrates quantum and classical computing to solve optimization problems, combining the strengths of both computational paradigms. Firstly, when the QAOA+ algorithm is used to solve the MIS problem on a subgraph, the classical computer is responsible for optimizing the parameters, while the quantum computer calculates the expected function value. This creates a feedback loop where classical and quantum components collaborate, with the classical computer adjusting parameters based on the output of the quantum computation. Secondly, the classical computer generates a new subgraph based on the structural information of the target graph. The quantum computer then solves the MIS problem on this subgraph and communicates the optimization results to the classical computer, which decides whether to expand the subgraph further based on the feedback. This forms another layer of collaboration between the classical and quantum components. In summary, PQA harnesses the superposition and parallelism of the quantum computer to solve the MIS problem on each subgraph. It collaborates with the classical computer, which handles tasks such as obtaining target graph information and adjusting parameters. This approach ensures that both quantum and classical components share the resource overheads rather than having one handle the entire burden alone.

\section{Numerical simulations}\label{simulation}
In our simulations, we compare the performance among various algorithms on three graph types, including ER graphs with edge probabilities of 0.4 and 0.5, as well as 3-regular graphs. There are 20 randomly generated graphs in each type, with 14 nodes (i.e., $n = 14$). For PQA, the tolerance on function values is set to $\xi = 0.1$, and the quantum state $|s\rangle$ is initialized to $| \bold 0 \rangle$ at the start of each optimization round. For both DS-QAOA+ and QLS, the initial quantum state is set to $| \bold 0 \rangle$ at the beginning of each optimization run. For QLS, the local subgraph is determined by a root node, $n_{\text{root}}$, and a free parameter, $N_{s}$. Specifically, the subgraph consists of the root node and all nodes whose distance from the root is $\le N_{s}$, along with all relevant edges among these nodes. Here, the root node is randomly selected, and the free parameter $N_s$ is the minimum value in the list $l_{\text{max}}$. The list $l_{\text{max}}$ stores the maximum shortest distances from any node to all other nodes in the original graph. For the three algorithms, the classical outer parameter optimization loop ends when the convergence tolerance \cite{MLI} satisfies $\varepsilon \le 0.001$. Additional details on the comparisons are provided below.

\subsection{The AR obtained by various algorithms with the same number of optimization runs and level depths}\label{AR_comparison}

To compare the performance of various quantum algorithms, we present the optimal AR (OAR) and average AR (AAR) achieved by each algorithm on different types of graphs, using the same number of runs and level depths. Concretely, we set the level depth from one to five, and each algorithm is executed with $100$ optimization runs at each level depth. For each graph type, this procedure yields 20 independent OAR (and AAR) values. The mean OAR (AAR) is then calculated as the average of these 20 values. Additionally, to establish a baseline for comparison, we include the AR obtained by the hill-climbing algorithm (i.e., a classical local search algorithm with wide applications)  \cite{hill_climbing} and a classical counterpart to PQA, which follows the same graph construction and expansion strategies described in Section~\ref{construction_rules}. However, instead of employing QAOA+ to approximate the MIS solution for each subgraph, we solve each subproblem using the hill-climbing algorithm. Notably, when solving each subproblem using the hill-climbing algorithm, the initial solution is inherited from the approximate solution obtained from the previous subgraph, except for the first subgraph, where an initial feasible solution is randomly generated.

\begin{figure*}[htbp]
	\centering
	\setlength{\abovecaptionskip}{0.cm}
	\subfigure[OAR, ER graphs with prob = 0.4]{	\includegraphics[width=0.3\textwidth]{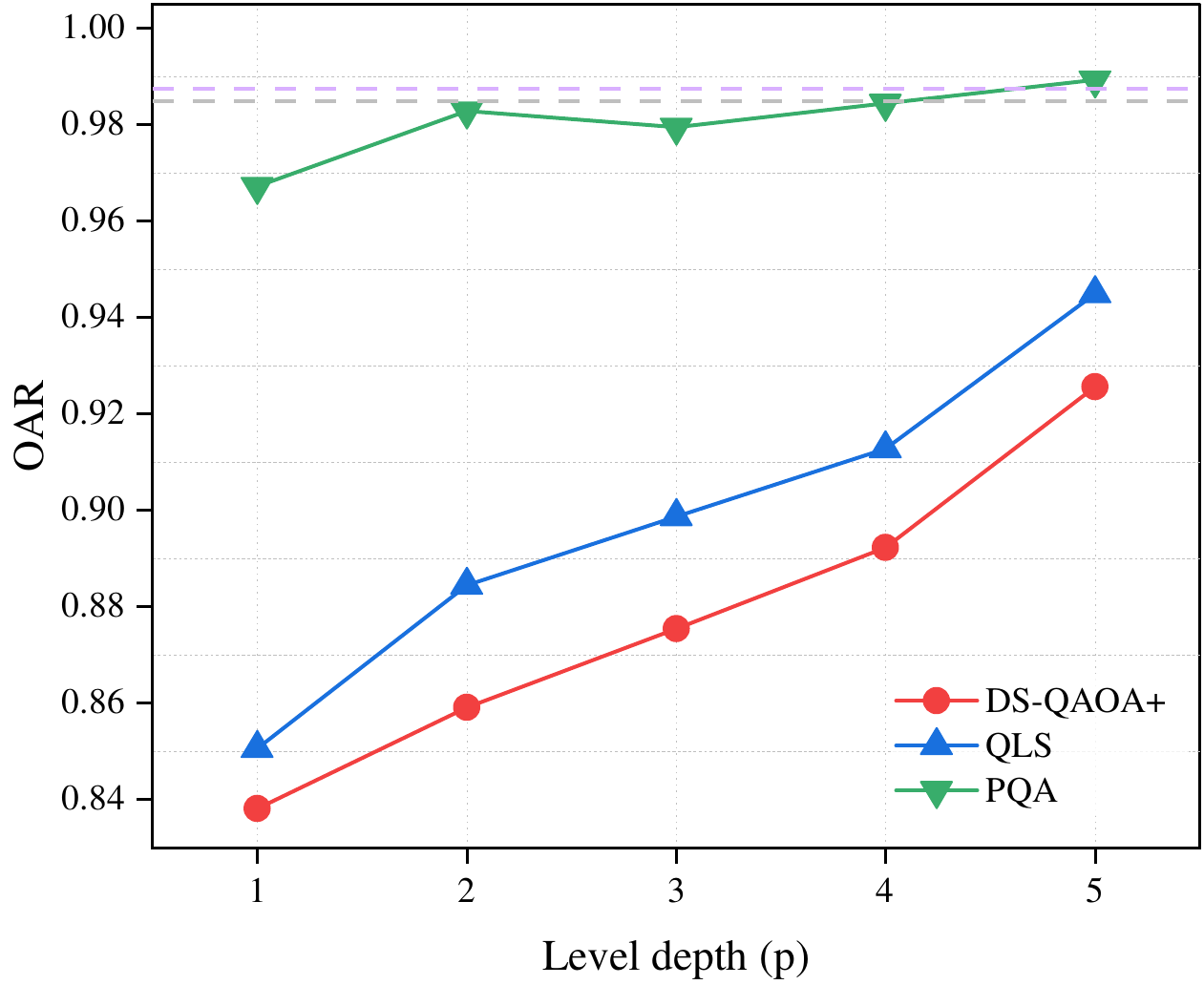}}
	\subfigure[ OAR, ER graphs with prob = 0.5]{
		\includegraphics[width=0.3\textwidth]{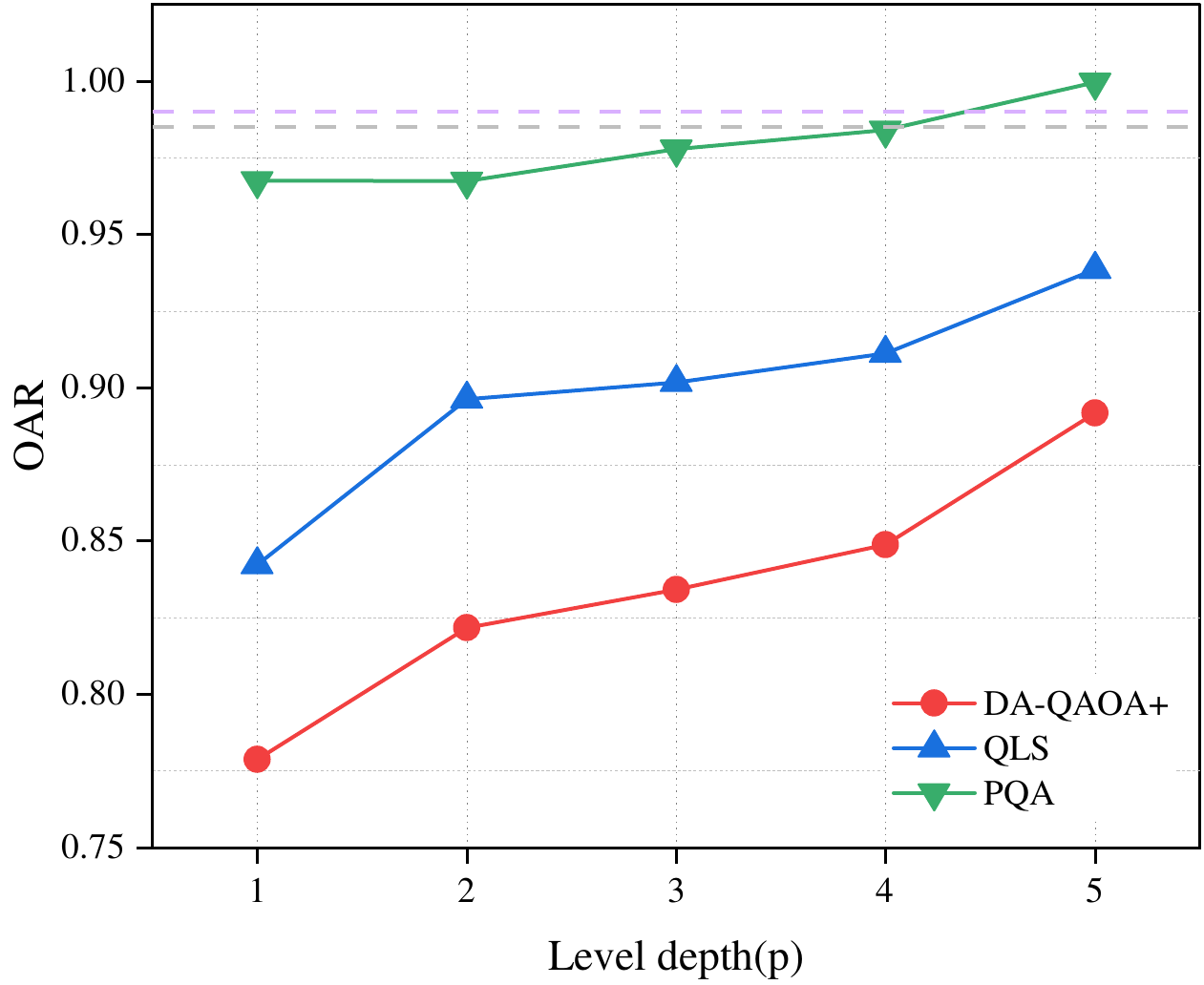}}
        \subfigure[ OAR, 3-regular graphs]{
		\includegraphics[width=0.3\textwidth]{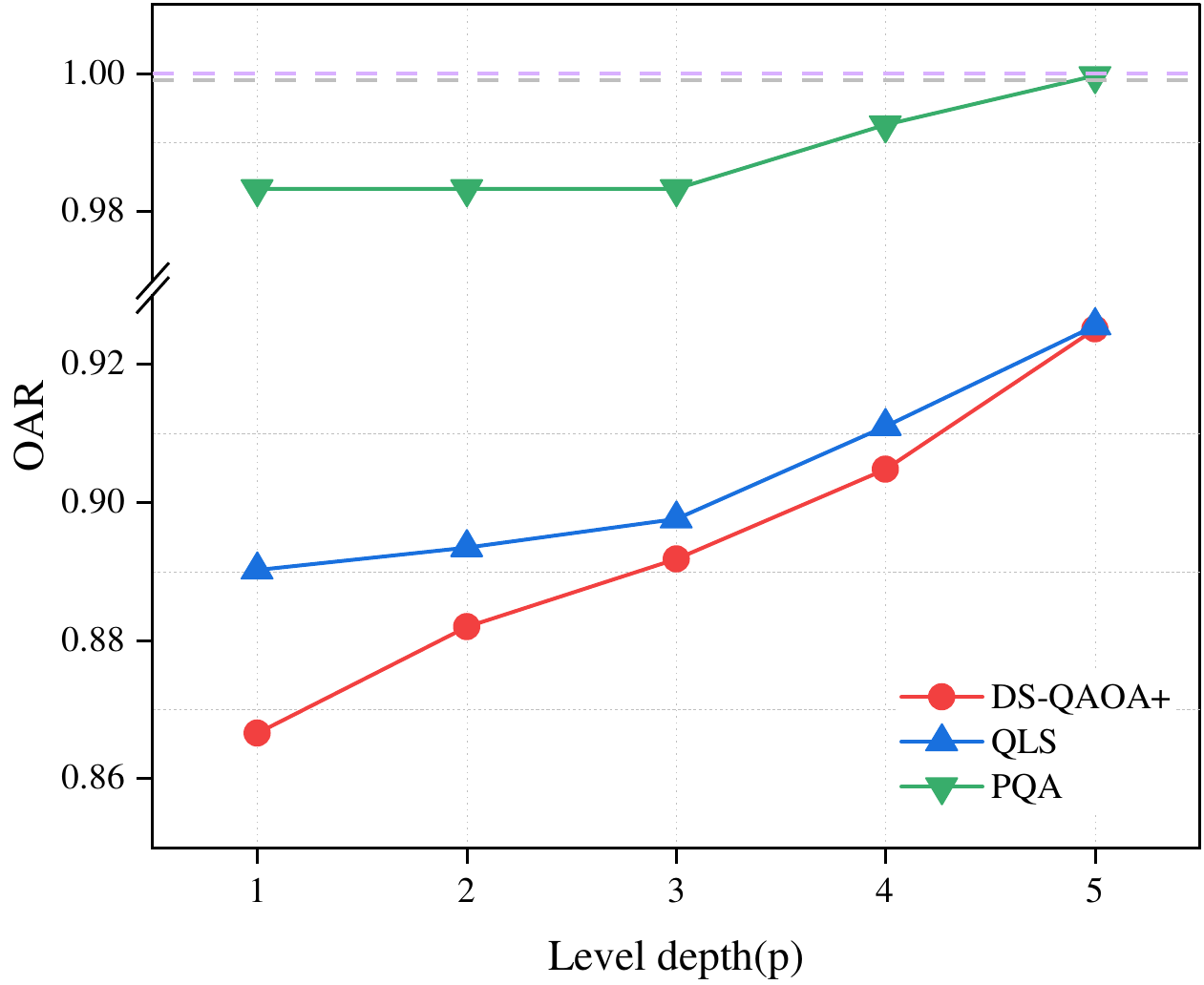}}

        \subfigure[ AAR, ER graphs with prob = 0.4]{	\includegraphics[width=0.3\textwidth]{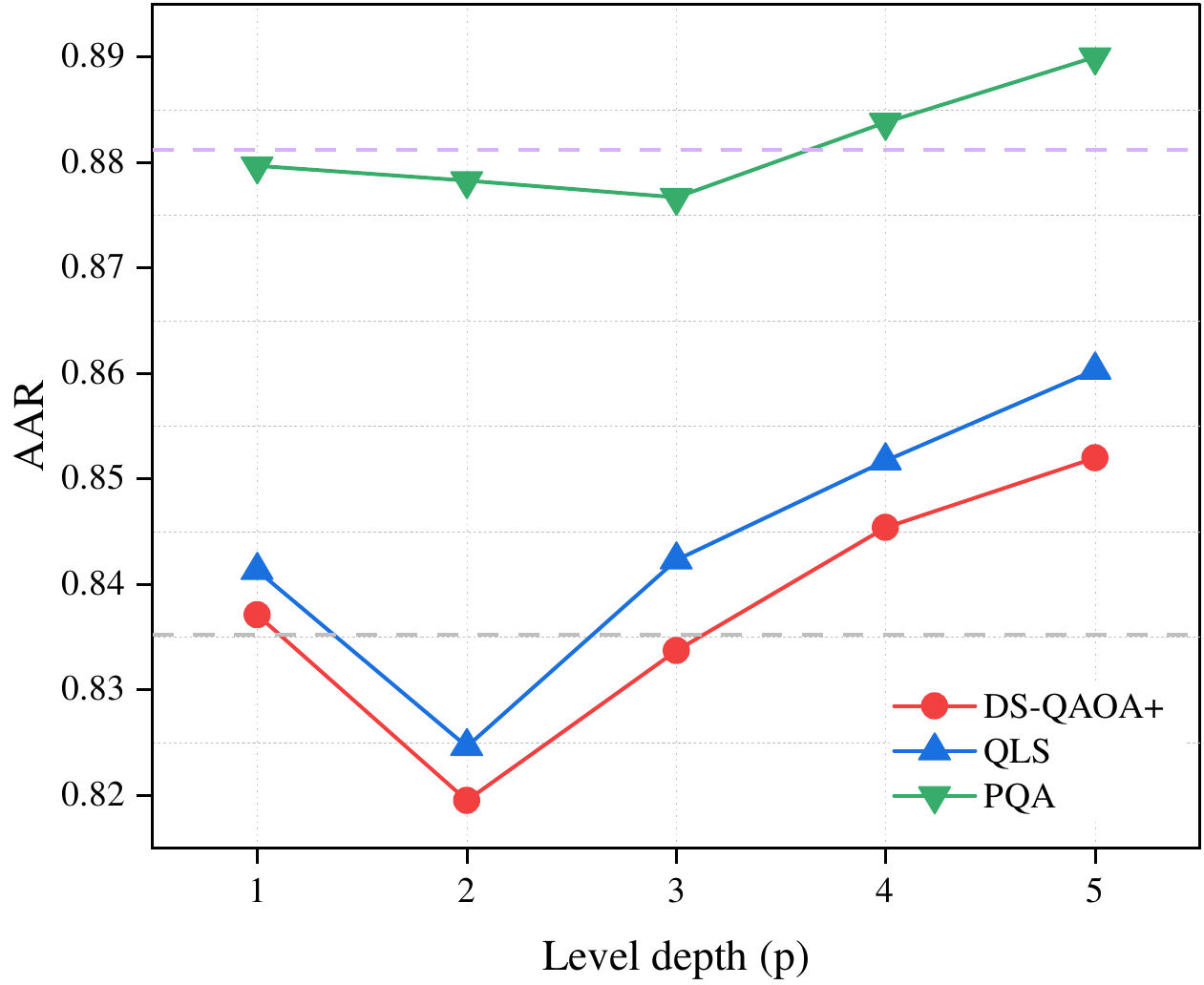}}
	\subfigure[ AAR, ER graphs with prob = 0.5]{
		\includegraphics[width=0.3\textwidth]{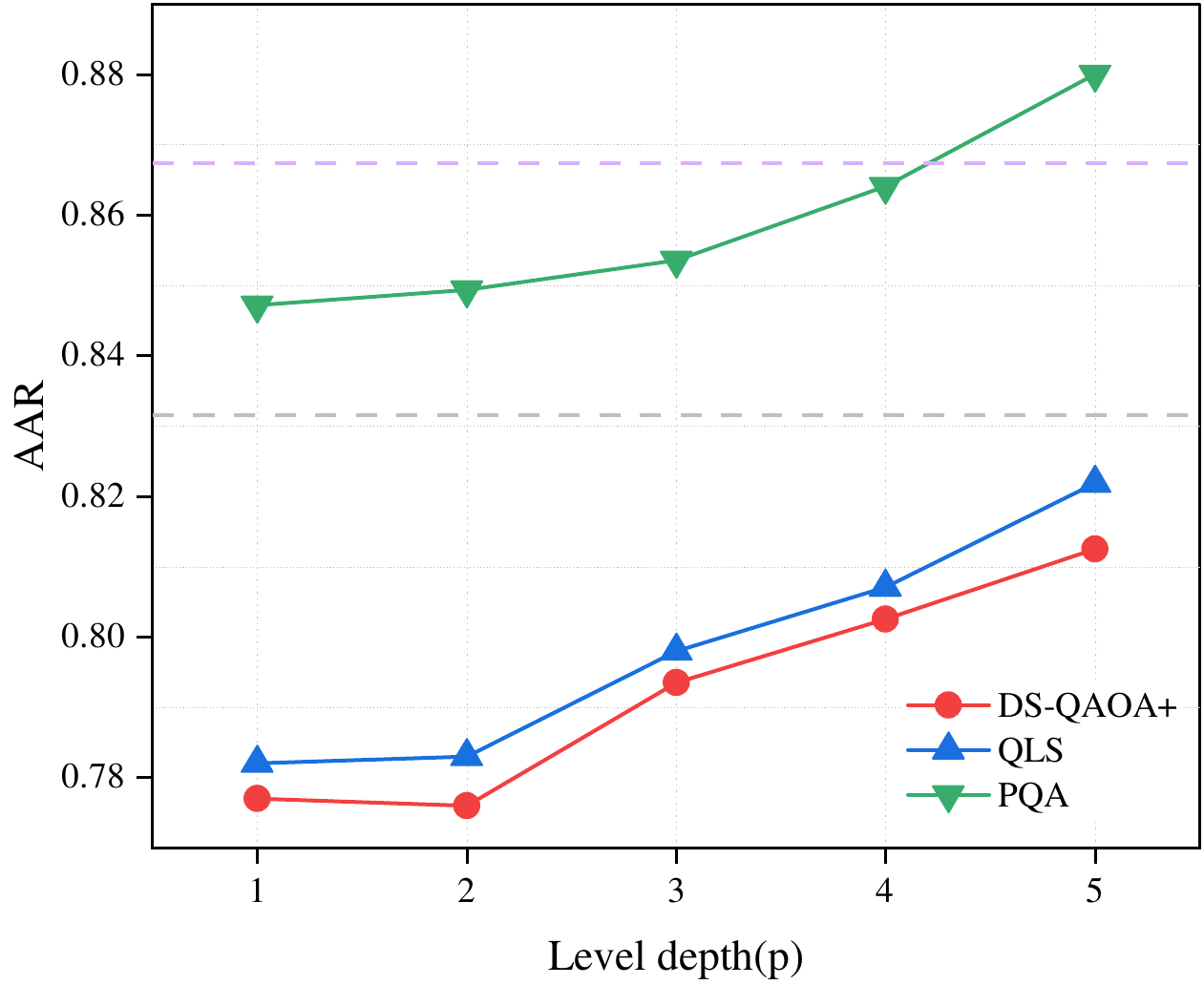}}
	\subfigure[ AAR, 3-regular graphs]{
		\includegraphics[width=0.3\textwidth]{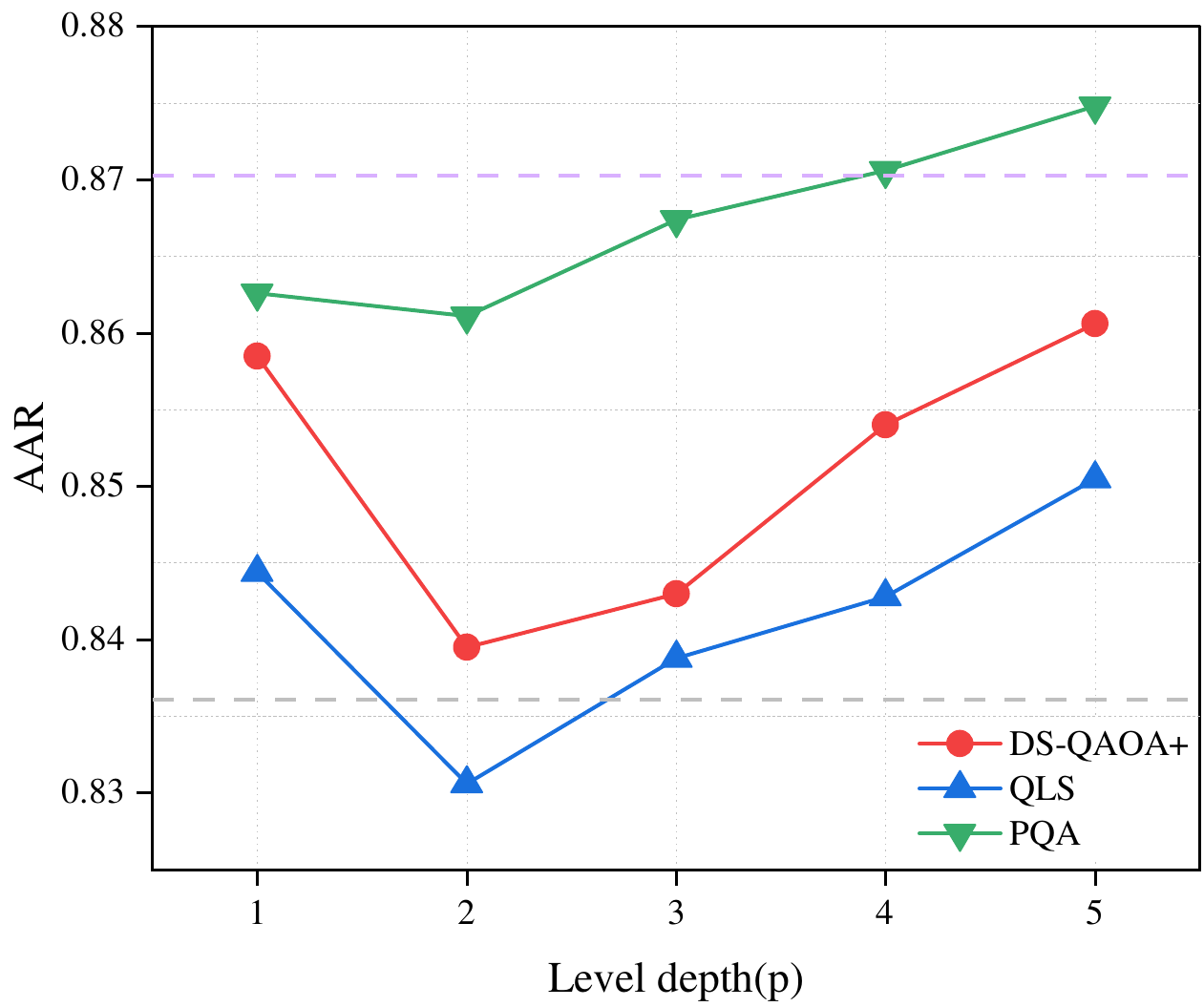}}
	\caption{The AR achieved by various algorithms. The gray dotted line corresponds to the AR obtained by the hill climbing algorithm, and the purple dotted line corresponds to the AR obtained by the classical counterpart of PQA. In each subplot, the curves depict the AR achieved by different algorithms across various graph types with increasing level depth $p$.}	\label{AR_comparison_fig}
\end{figure*}

\medskip
The OAR is defined as the maximum approximation ratio obtained over multiple optimization runs. It represents the best performance of the algorithm and provides a measure of the highest-quality solution achieved in the experiments. The AAR is the average of the approximation ratios obtained in multiple optimization runs, providing a measure of stability. From these numerical results illustrated in Figure~\ref{AR_comparison_fig}, we observe a clear trend that the achieved OAR increases as the level depth grows, indicating that the solutions obtained are closer to the global optimal solution with reduced approximation error. This improvement is primarily due to the increased ansatz depth, which enhances the expressive power of the quantum circuit, allowing it to represent a broader range of quantum states and enable a more effective search for high-quality solutions within the Hilbert space. \textbf{Notably, the results also show that, given the same number of runs and level depths, PQA achieves a higher OAR and a significantly higher AAR compared with other quantum algorithms. This suggests that PQA not only excels in finding high-quality solutions but also maintains a more stable performance across multiple optimization runs. The higher AAR further implies that the solutions of PQA exhibit lower variance, ensuring that it reliably produces quasi-optimal solutions regardless of random parameter initialization. Such stability is crucial for practical applications as it reduces the risk of performance fluctuations, making PQA a more robust choice. The specific improvements are as follows.} 

\begin{itemize} 
    \item On ER graphs with an edge probability of 0.4, the OAR achieved by PQA is approximately $8.240\%$ and $13.668\%$ higher than that achieved by QLS and DS-QAOA+, respectively. The AAR achieved by PQA is approximately $3.766\%$ and $4.416\%$ higher than that achieved by QLS and DS-QAOA+, respectively.
    \item On ER graphs with an edge probability of 0.5, the OAR achieved by PQA is approximately $8.128\%$ and $14.422\%$ higher than that achieved by QLS and DS-QAOA+, respectively. The AAR achieved by PQA is approximately $6.048\%$ and $6.658\%$ higher than that achieved by QLS and DS-QAOA+, respectively.
    \item On 3-regular graphs, the OAR obtained by PQA is approximately $8.486\%$ and $9.438\%$ higher than that achieved by QLS and DS-QAOA+, respectively. The AAR obtained by PQA is approximately $2.588\%$ and $1.618\%$ higher than that achieved by QLS and DS-QAOA+, respectively.
\end{itemize}

\medskip
Compared with directly applying the hill-climbing algorithm to solve the original problem, the classical counterpart to PQA significantly improves the solution stability by progressively refining the obtained solution on each subgraph, demonstrating the effectiveness of our method in addressing the problem in an incremental manner. The results also show that the AR obtained by PQA  can be higher than the classical counterpart to PQA, which benefits from quantum superposition and quantum parallelism. Specifically, the classical local search algorithm typically requires multiple verification and correction steps to guarantee solution feasibility  \cite{QLS}, and the feasible solution space constructed by the classical local search algorithm is inherently limited due to the sequential and local nature of its search process, as it evaluates and updates only a single solution per iteration. This limitation is mitigated in PQA, as it constructs a larger feasible solution space through quantum superposition and elaborate circuit designs. By leveraging quantum parallelism, QAOA+ evaluates multiple feasible solutions in parallel. Therefore, in PQA, QAOA+ is preferred over classical heuristic algorithms for solving the MIS problem on each subgraph.

\subsection{The quantum resource consumption of various algorithms when achieving the same AR} \label{resource_comparison}

In the previous subsection, we compared the obtained OAR and AAR of different algorithms given the same number of runs and level depths. However, comparing algorithms solely based on the same number of runs has limitations, as some algorithms may achieve a similar OAR with fewer runs, which could lead to an unfair assessment of their efficiency. To provide a more balanced evaluation, we further compare the total resource consumption of each algorithm when achieving the same given AR. We selected several key metrics to evaluate performance, including the number of iterations, the number of qubits, the number of multi-qubit controlled $R_x$ gates, and the algorithm runtime. By combining these metrics, we can more clearly and comprehensively reflect the resource consumption of different algorithms, offering valuable insights for performance optimization in practical applications. In Appendix~\ref{calculate_resources}, we give the methods of calculating the total resources for each algorithm.

\begin{figure*}[htbp]
	\centering
	\setlength{\abovecaptionskip}{0.cm}

	\subfigure[Total iterations, ER with prob = 0.4]{
		\includegraphics[width=0.3\textwidth]{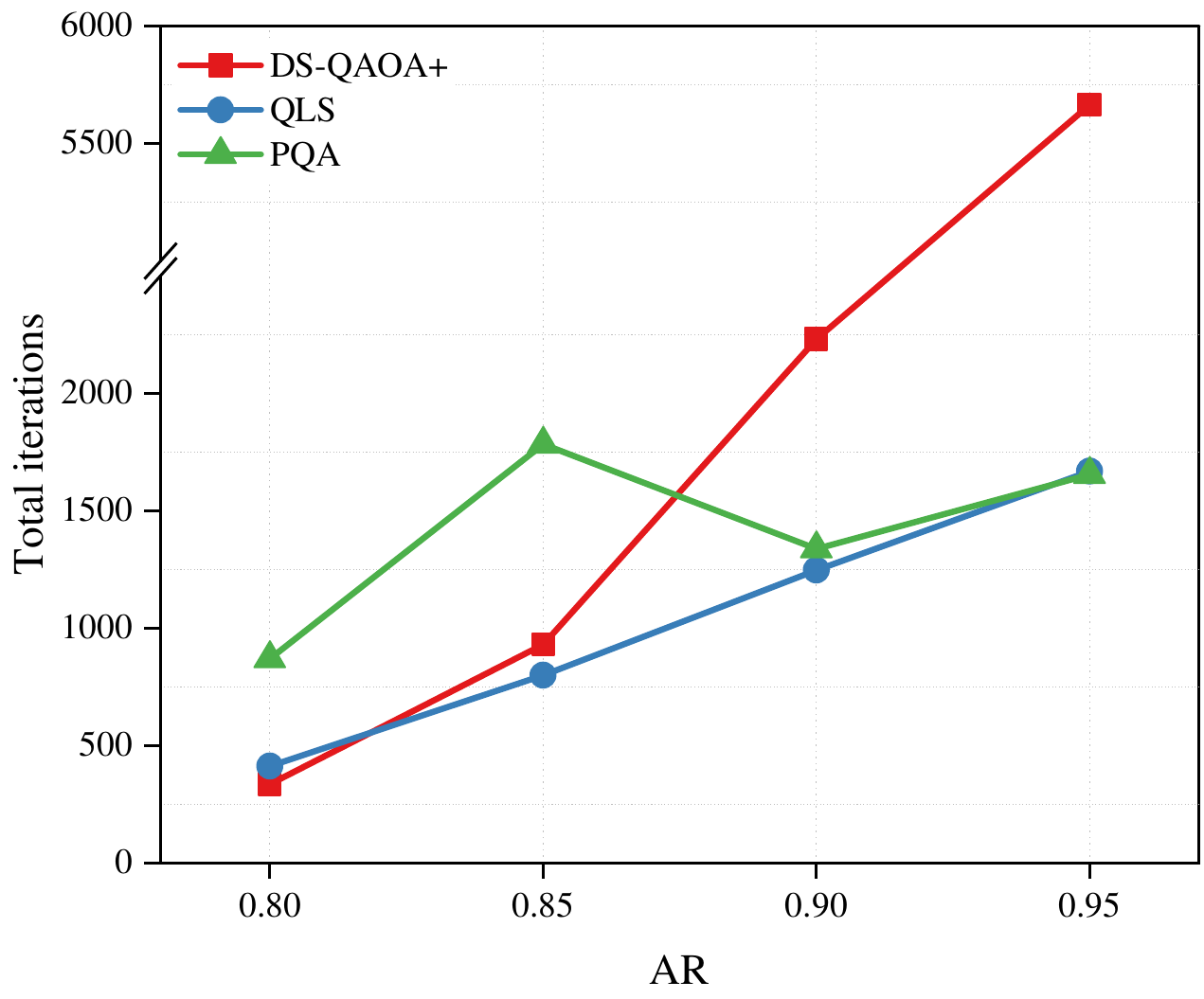}}
	\subfigure[ Total iterations, ER with prob = 0.5]{
		\includegraphics[width=0.3\textwidth]{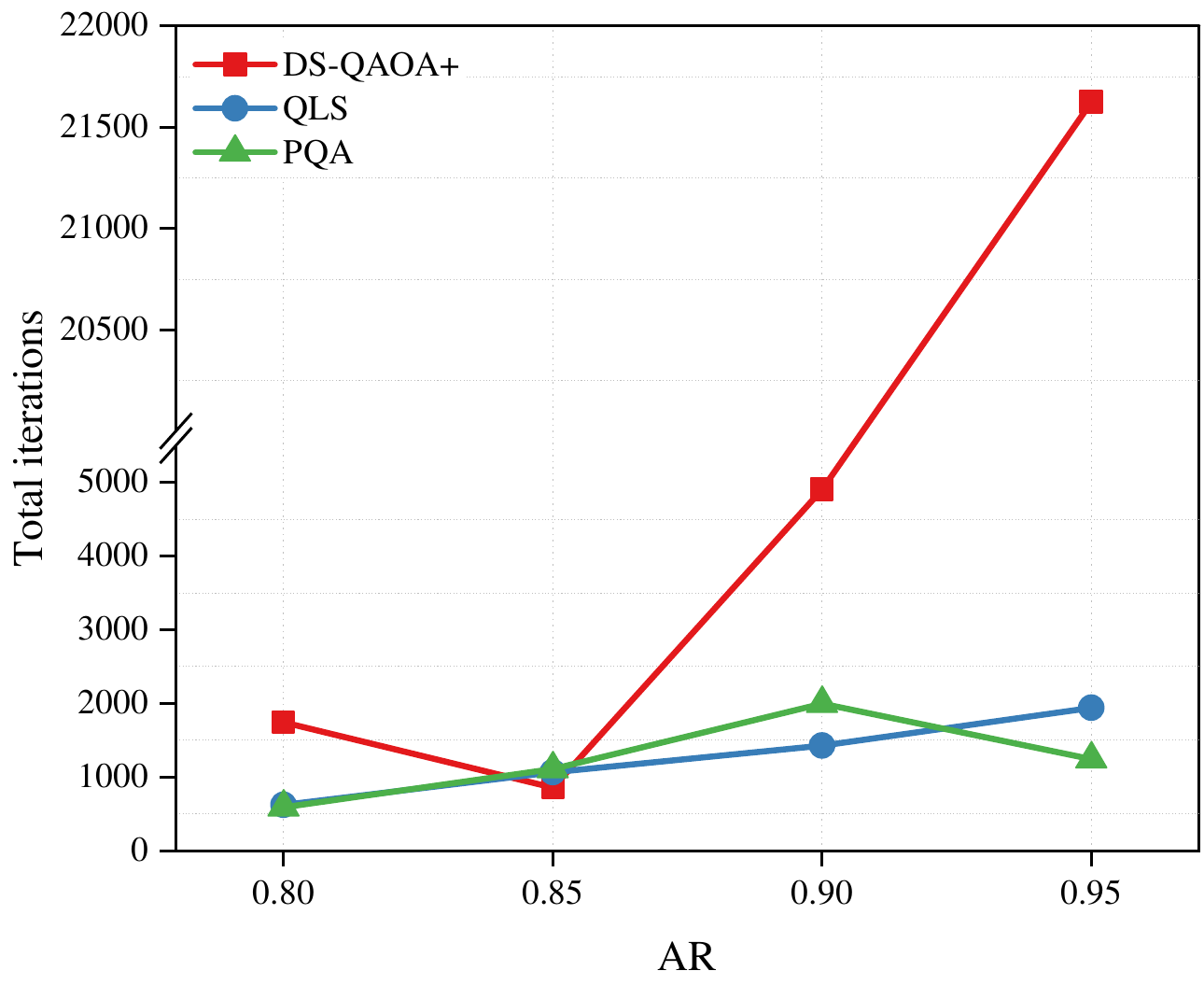}}
        \subfigure[ Total iterations, 3-regular graphs]{
		\includegraphics[width=0.3\textwidth]{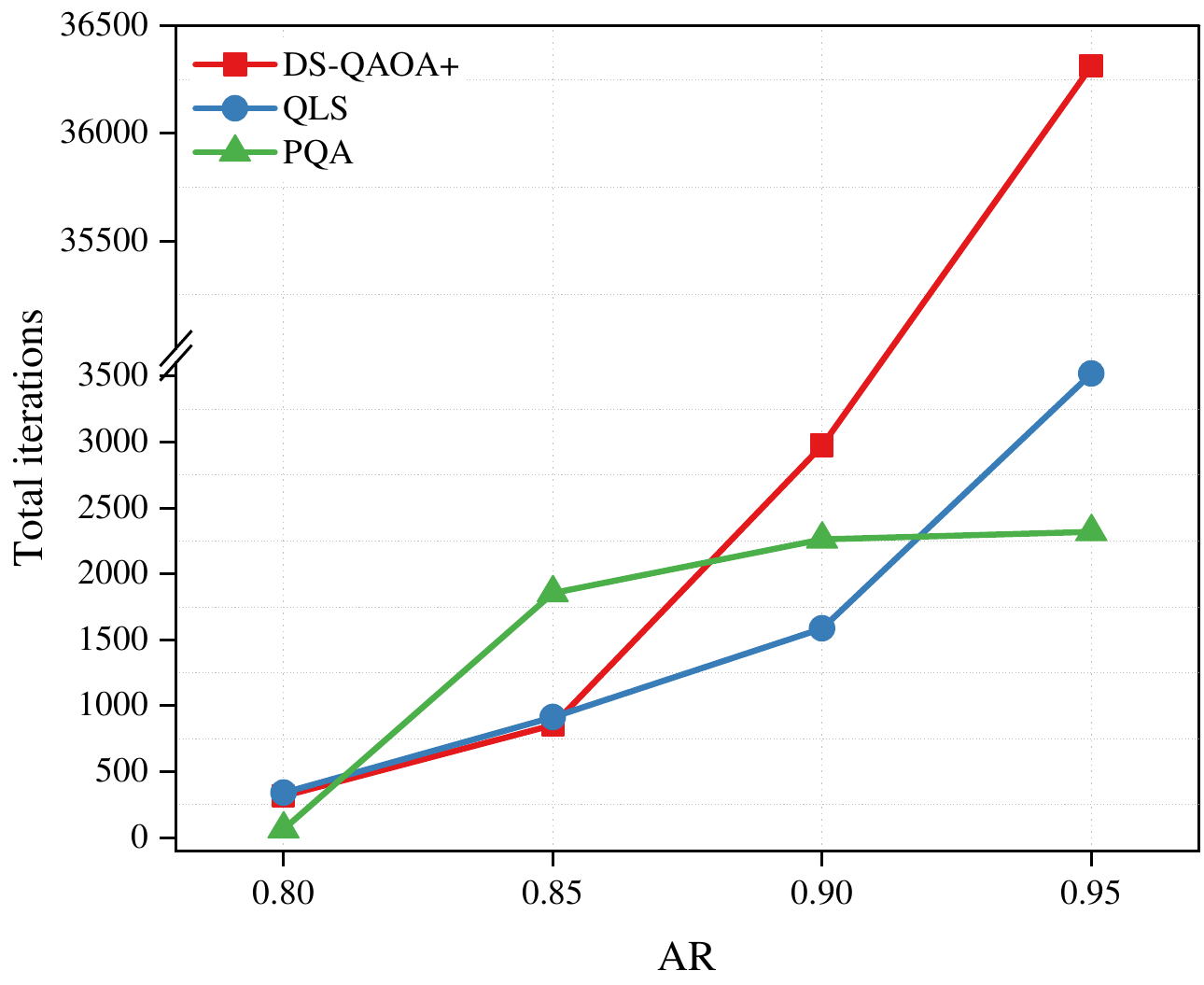}}

        \subfigure[ Total runtime, ER with prob = 0.4]{
		\includegraphics[width=0.3\textwidth]{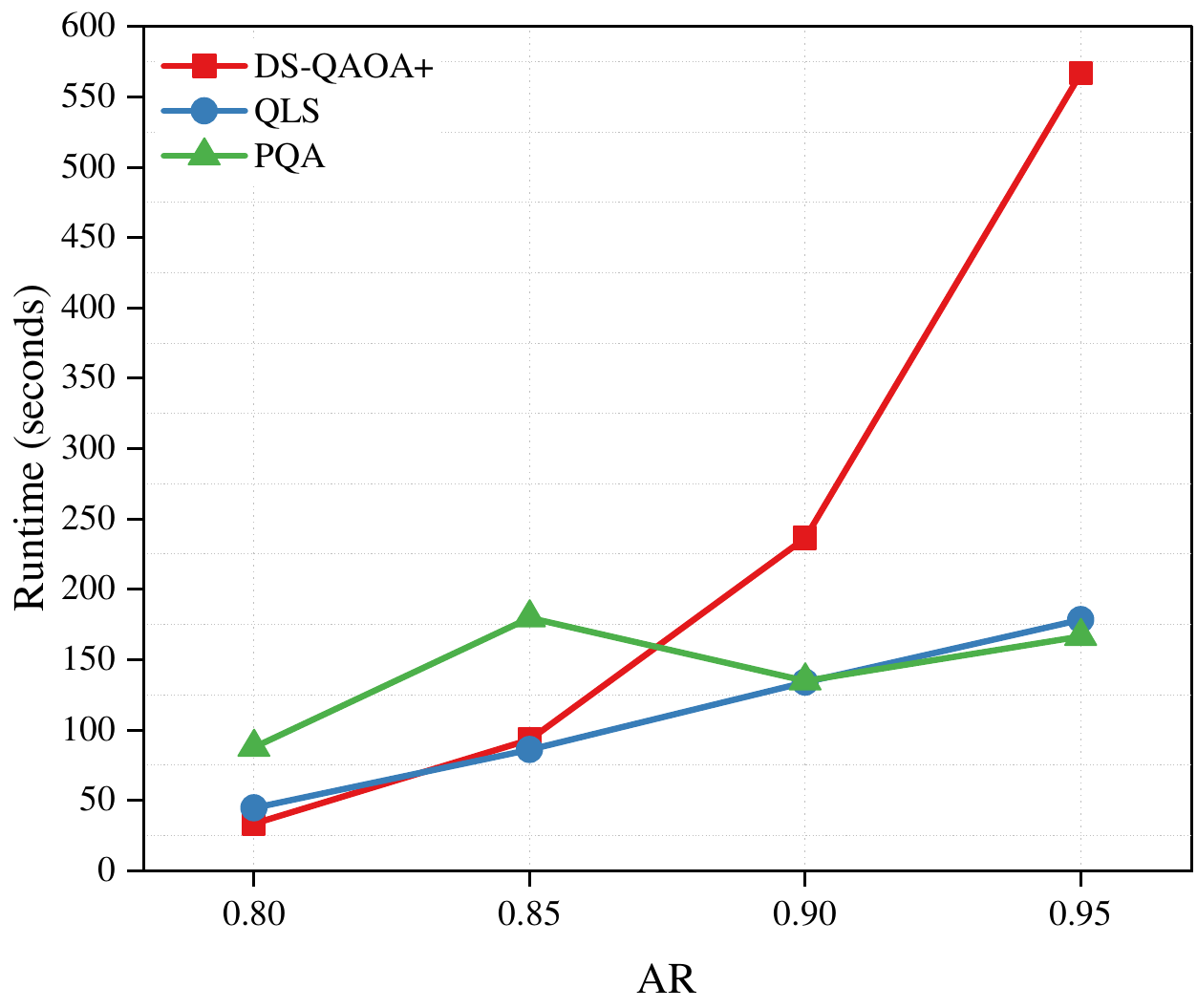}
	}
	\subfigure[ Total runtime, ER with prob = 0.5]{
		\includegraphics[width=0.3\textwidth]{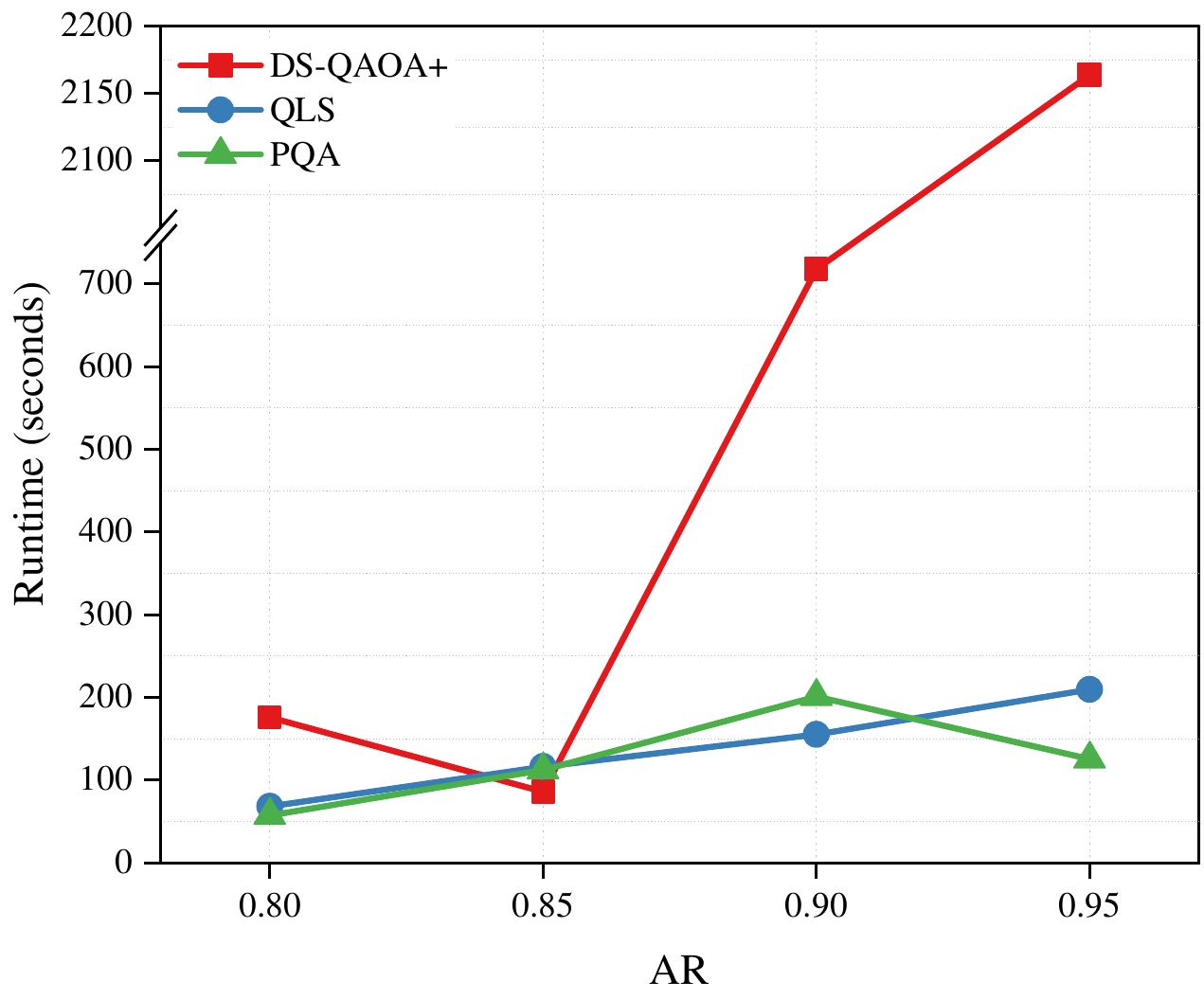}}
	\subfigure[ Total runtime, 3-regular graphs]{
		\includegraphics[width=0.3\textwidth]{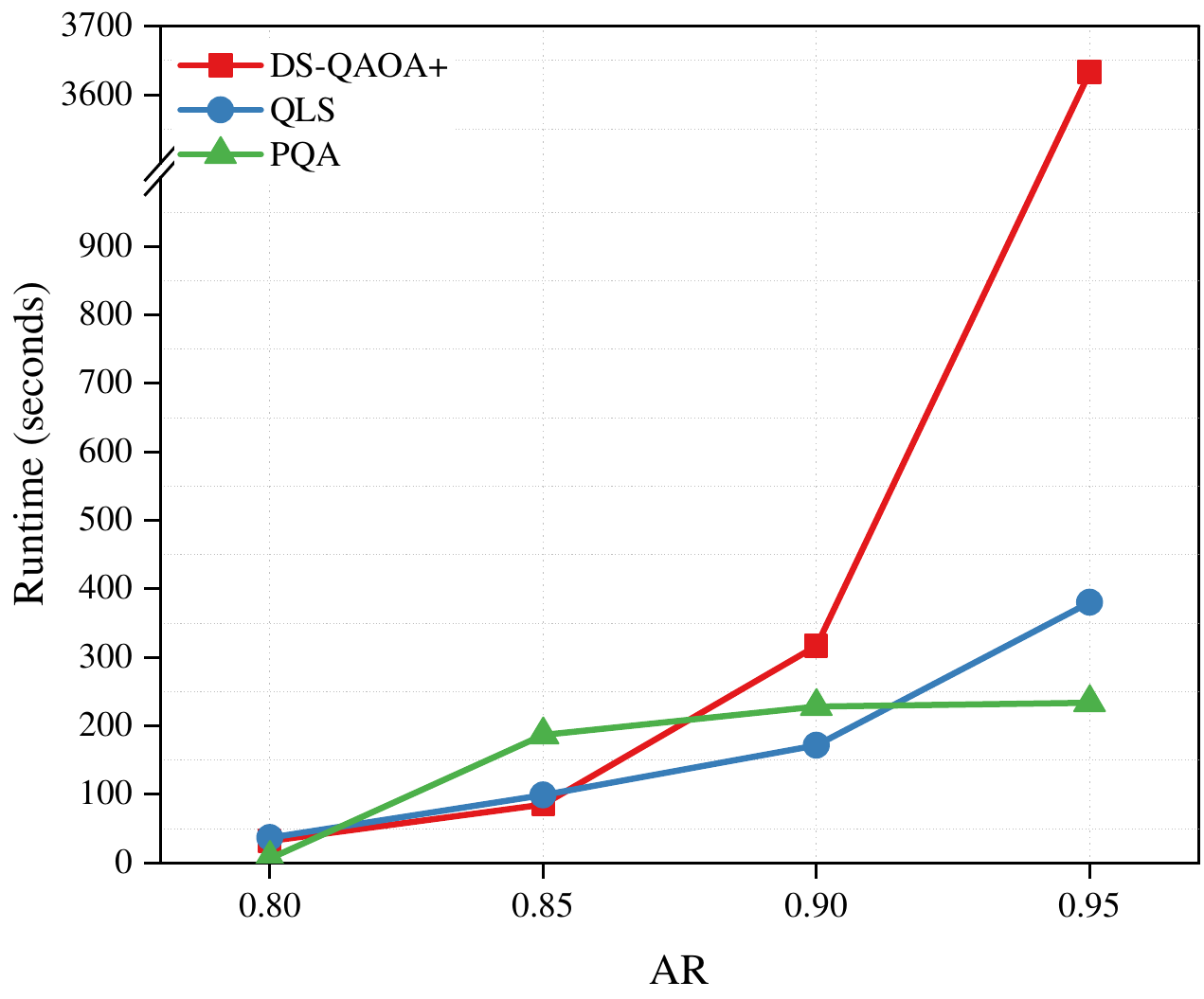}}

        \subfigure[ Total qubits, ER with prob = 0.4]{
		\includegraphics[width=0.3\textwidth]{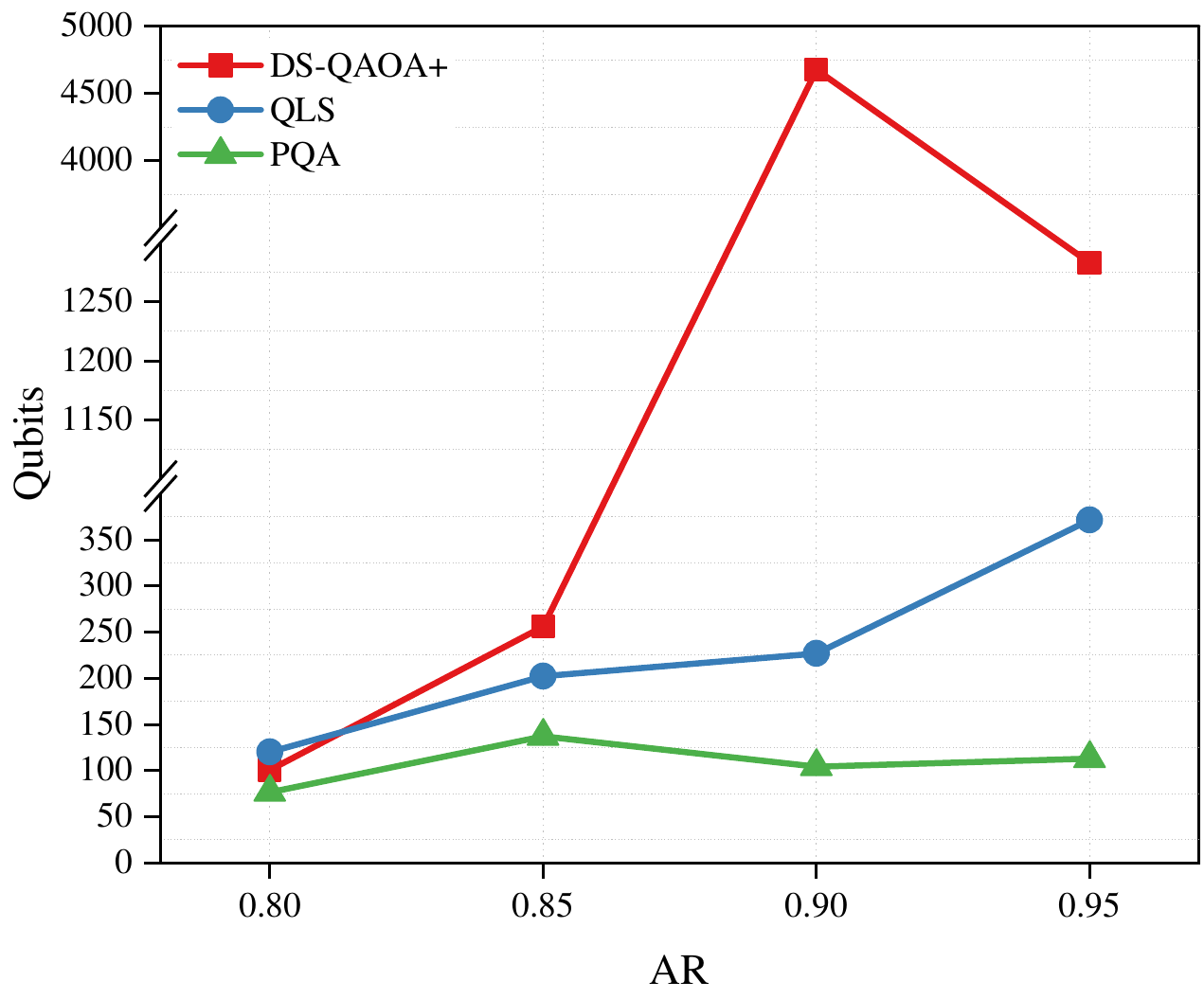}
	}
	\subfigure[ Total qubits, ER with prob = 0.5]{
		\includegraphics[width=0.3\textwidth]{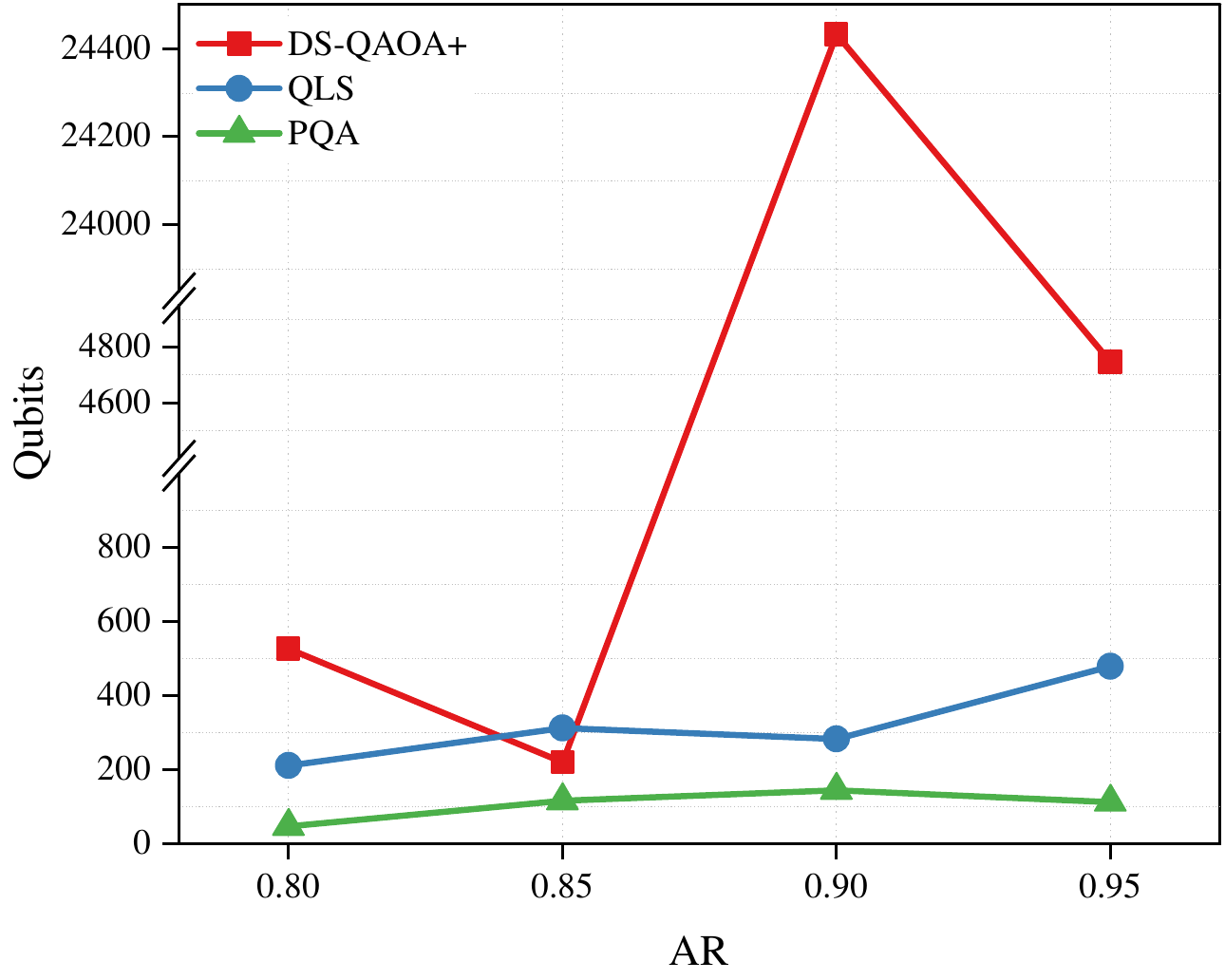}}
	\subfigure[ Total qubits, 3-regular graphs]{
		\includegraphics[width=0.3\textwidth]{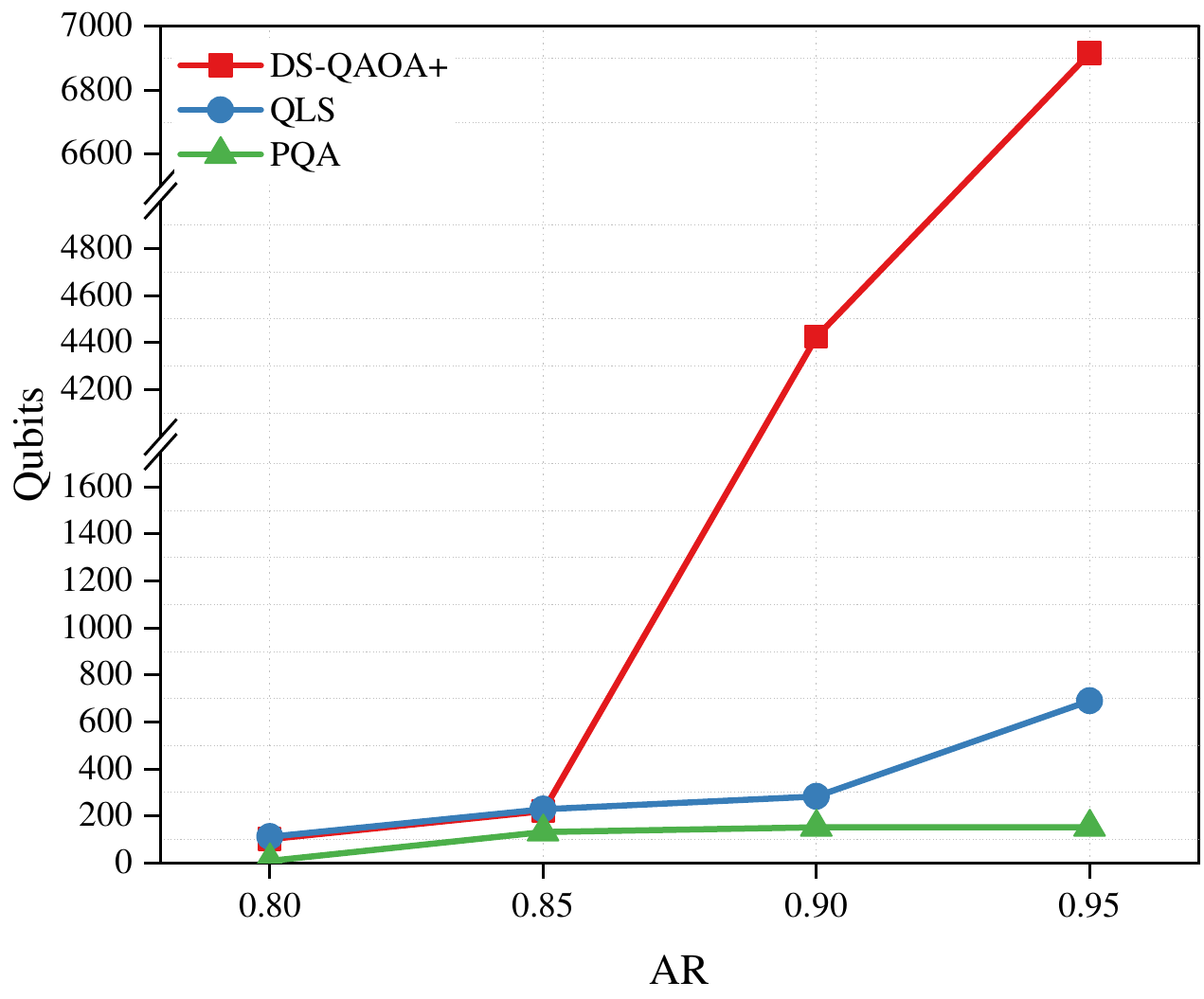}}

        \subfigure[ Total multi-qubit gates, ER with prob = 0.4]{
		\includegraphics[width=0.3\textwidth]{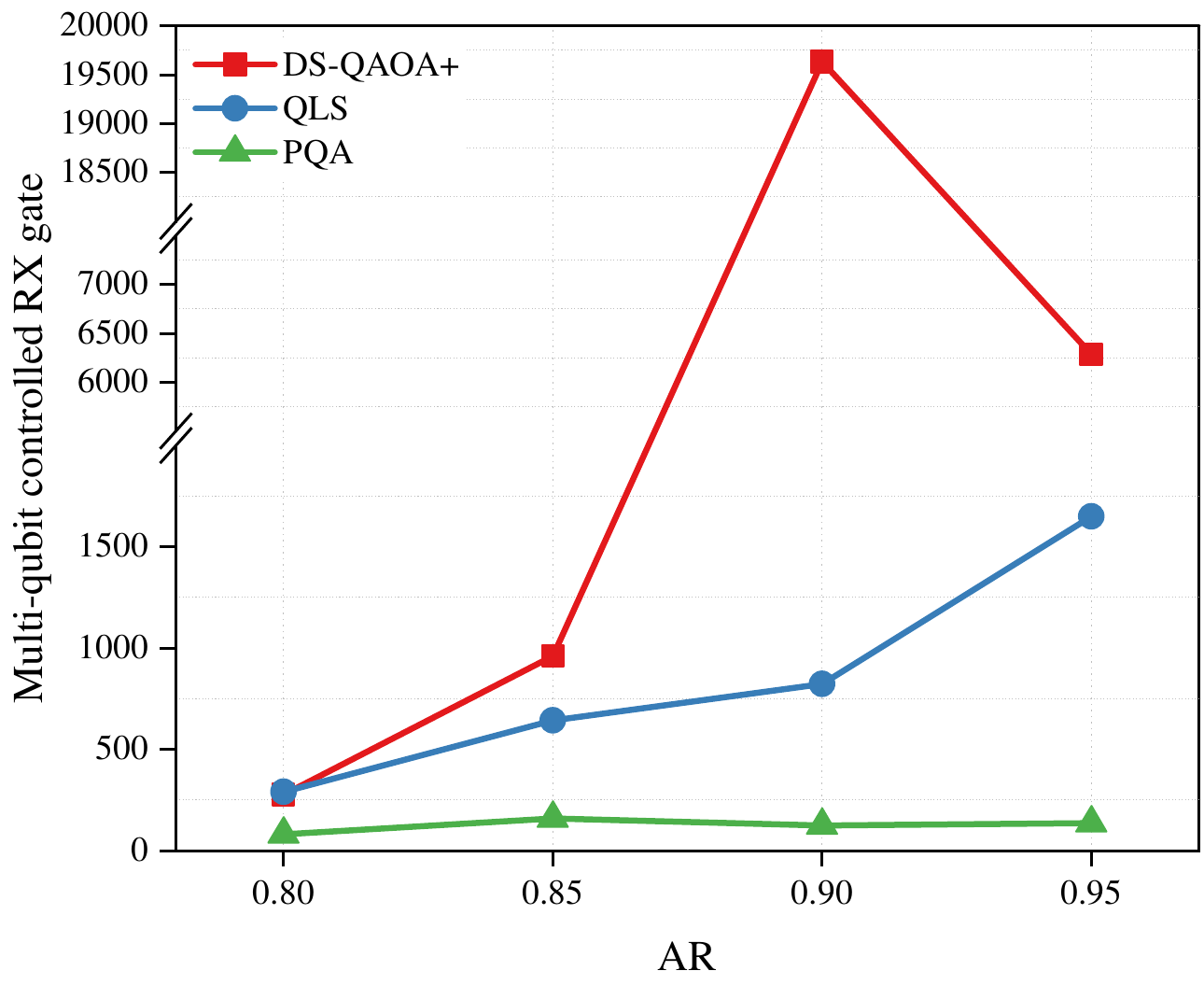}
	}
	\subfigure[ Total multi-qubit gates, ER with prob = 0.5]{
		\includegraphics[width=0.3\textwidth]{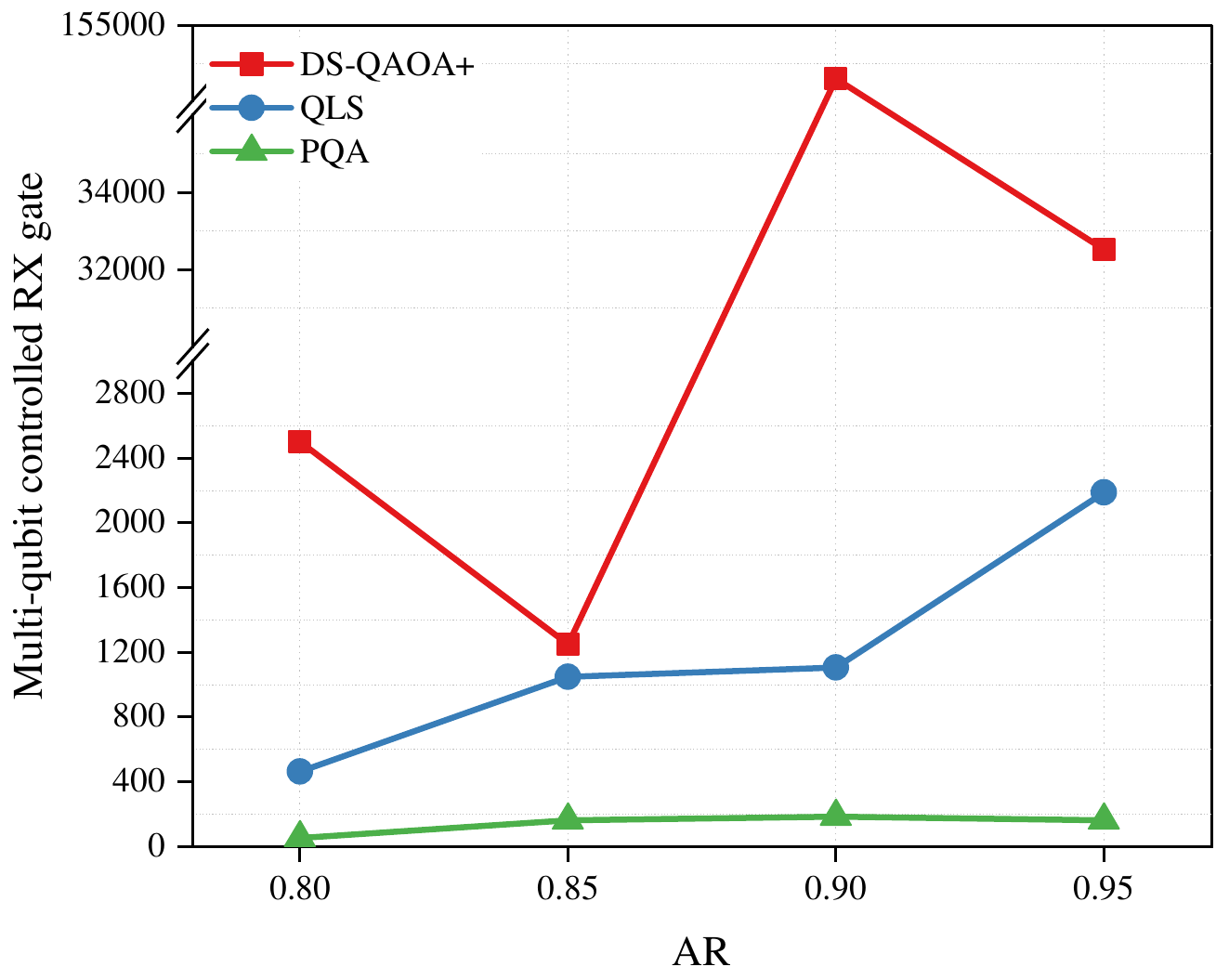}}
	\subfigure[ Total multi-qubit gates, 3-regular graphs]{
		\includegraphics[width=0.3\textwidth]{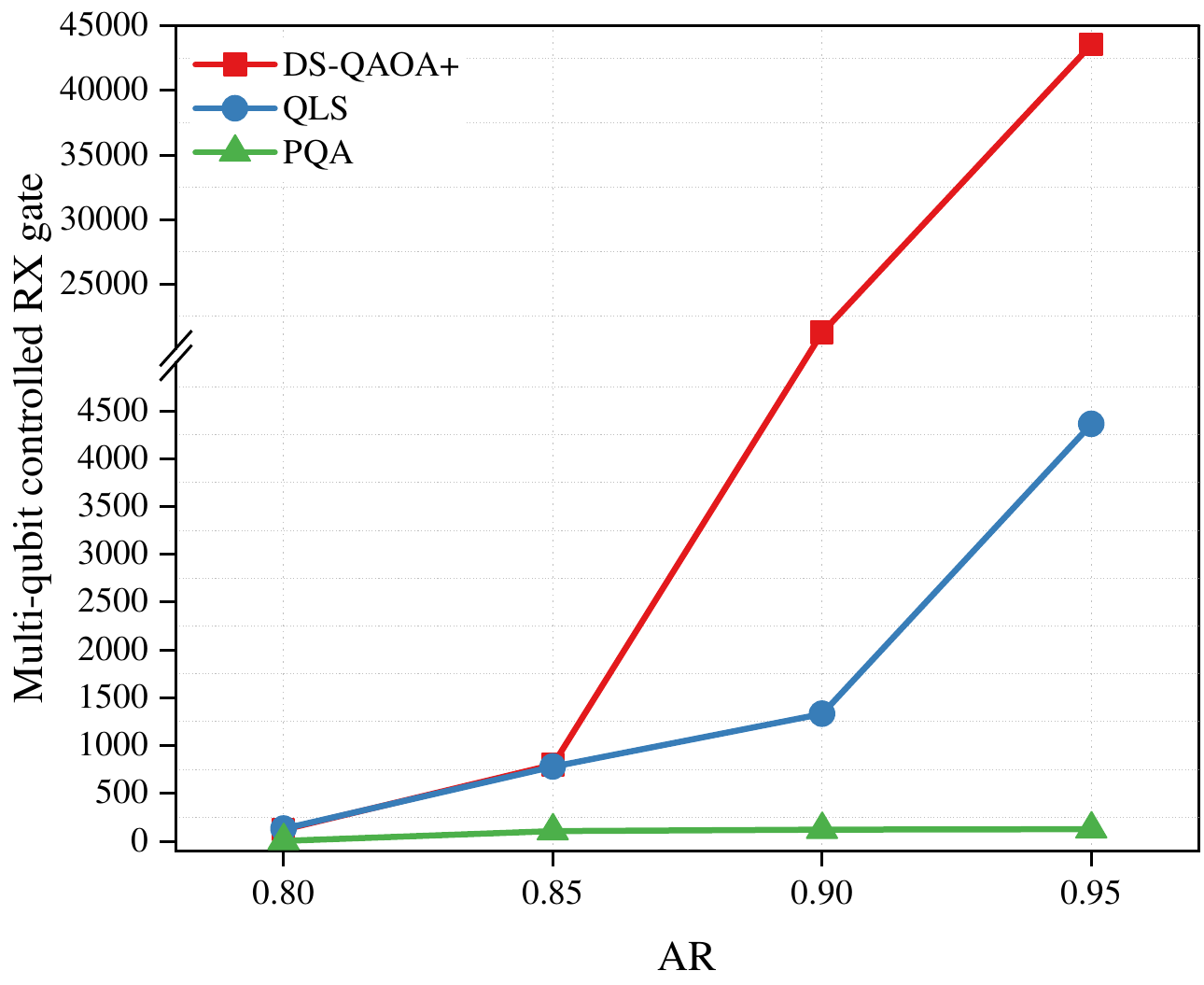}}
        
	\caption{The total resource consumption of various algorithms on different graph types.}	\label{resource_consumption}
\end{figure*}

\medskip
FIG.~\ref{resource_consumption} presents the total resource consumption required by different algorithms to achieve a given AR across various graph types. From the numerical results, we observe that the total resource consumption increases as the target AR increases, which aligns with expectations. This is because a higher AR requires the output solutions to be closer to the true solution, necessitating more optimization runs or deeper level depths to search for a better solution, increasing the overall resource consumption. However, DS-QAOA+ and PQA exhibit counterintuitive behavior in the resource consumption trends in some cases. Especially for DS-QAOA+, the total number of qubits and multi-qubit controlled $R_x$ gates for achieving AR = 0.9 is significantly higher than the corresponding consumption for achieving AR of 0.95. Upon examining the experimental data, we found that DS-QAOA+ reaches the target AR using fewer level depths but consumes more runs at an AR of 0.9. In contrast, at an AR of 0.95, DS-QAOA+ increases the number of level depths to achieve the given AR but reduces the number of runs required, as shown in FIG.~\ref{runs_consumption}. This phenomenon is related to the expressive ability of the PQC, which impacts the ability of the algorithm to find quasi-optima. Specifically, shallow level depths have limited expressive power and can explore only a smaller solution space during each optimization process. Therefore, to approach the target solution, the algorithm requires more runs to compensate for the limited search space explored. In contrast, deep level depths, with stronger expressive capabilities, can explore a broader solution space. This enables the algorithm to more precisely approach the target solution in each run,  thus achieving the same AR with fewer runs. However, deeper level depths are not always better, the greater the level depth means the greater the demand for quantum gate and circuit depth, which may limit the scale of problem-solving on the current NISQ device. In our calculations, the total consumption is determined by the product of the total number of runs and the average resource consumption per run. While increasing level depth inevitably raises the number of gates per run, the increase in gate count is relatively small compared with the difference in the number of optimization runs between AR of 0.9 and AR of 0.95. Moreover, increasing the level depth does not lead to an increase in the number of qubits consumed per run. As a result, for AR = 0.9, the total number of multi-qubit controlled $R_{x}$ gates and qubits is higher compared with AR of 0.95. The above analysis can also explain some counterintuitive phenomena in PQA.

\medskip
From the numerical simulation results in FIG.~\ref{resource_consumption}, we also observed that for smaller target AR (e.g., 0.8 and 0.85), DS-QAOA+ and QLS typically achieve the given AR with fewer iterations and shorter runtime compared with PQA. As shown in FIG.~\ref{runs_consumption}, for AR values of 0.8 and 0.85, the total number of optimization runs consumed by DS-QAOA+, QLS, and PQA is similar. However, since PQA performs more optimization rounds per run than DS-QAOA+ and QLS, it results in a higher total iteration consumption. Moreover, the consumption of iterations has a significant impact on the overall runtime as shown in Appendix~\ref{calculate_runtime}, with more iterations generally leading to higher runtime consumption. Nevertheless, it is important to note that PQA consumes significantly fewer qubits and multi-qubit controlled gates compared with DS-QAOA+ and QLS. While the number of iterations is important, the number of qubits and gates required per run more directly determines the problem size that can be supported by resource-constrained quantum devices. Excessive qubit and gate demands per iteration may limit the ability of hardware to solve larger-scale problems efficiently. Therefore, sacrificing some iterations to reduce qubit and gate consumption is a justified trade-off, as it allows the algorithm to tackle larger problems within the constraints of current quantum hardware. However, this does not imply that PQA always requires the most iterations for any AR. Numerical results show that as the target AR increases, the efficiency of PQA becomes more evident. This is mainly because PQA can achieve the desired AR with fewer runs than DS-QAOA+ and QLS, significantly reducing overall resource consumption. By minimizing the quantum resources required while achieving high performance, PQA may be increasingly promising. Detailedly, for each graph type, the resource consumption of PQA compared with DS-QAOA+ and QLS for a given AR of 0.95 is as follows.
\begin{itemize}
    \item On the ER graphs with an edge probability of 0.4, the number of iterations consumed by PQA is approximately $29.17\%$ and $99.16\%$ of those consumed by DS-QAOA+ and QLS, respectively. The amount of runtime consumed by PQA is approximately $29.38\%$ and $93.48\%$ of those consumed by DS-QAOA+ and QLS, respectively. The number of qubits consumed by PQA is approximately $8.79\%$ and $30.36\%$ of those consumed by DS-QAOA+ and QLS, respectively. The number of multi-qubit controlled gates consumed is $2.16\%$ and $8.24\%$ of those consumed by DS-QAOA+ and QLS, respectively.
    
    \item On the ER graphs with an edge probability of 0.5, the number of iterations consumed by PQA is approximately $5.70\%$ and $64.00\%$ of those consumed by DS-QAOA+ and QLS, respectively. The amount of runtime consumed by PQA is approximately $5.80\%$ and $59.95\%$ of those consumed by DS-QAOA+ and QLS, respectively. The number of qubits consumed by PQA is approximately $2.34\%$ and $23.18\%$ of those consumed by DS-QAOA+ and QLS, respectively. The number of multi-qubit controlled gates consumed is $0.49\%$ and $7.36\%$ of those consumed by DS-QAOA+ and QLS, respectively.
    
    \item On the 3-regular graphs, the number of iterations consumed by PQA is approximately $6.39\%$ and $65.92\%$ of those consumed by DS-QAOA+ and QLS, respectively. The amount of runtime consumed by PQA is approximately $6.43\%$ and $61.49\%$ of those consumed by DS-QAOA+ and QLS, respectively. The number of qubits consumed by PQA is approximately $2.17\%$ and $21.76\%$ of those consumed by DS-QAOA+ and QLS, respectively. The number of multi-qubit controlled gates consumed is $0.29\%$ and $2.88\%$ of those consumed by DS-QAOA+ and QLS, respectively.
\end{itemize}

\section{Conclusion and outlook}\label{conclusion}
In this paper, we propose a heuristic quantum algorithm termed PQA to reduce the number of qubits in the application of QAOA+ when solving the MIS problem. The goal of PQA is to progressively construct a subgraph that is likely to contain the MIS solution of the original graph and then get the approximate solution by solving the MIS problem on this subgraph.
To induce such a subgraph, PQA starts with a small-scale initial subgraph and progressively expands its graph size utilizing heuristic expansion rules. After each expansion, PQA solves the MIS problem on the current subgraph using QAOA+. In each run, PQA repeats the graph
expansion and solving process until a predefined stopping condition is reached. The numerical results demonstrate that PQA can achieve higher OARs and AARs than DS-QAOA+ and QLS under the same optimization runs and level depths, indicating that PQA gets higher solution quality and has greater stability. Additionally, PQA is more efficient than DS-QAOA+ and QLS, which is reflected that PQA consumes fewer qubits and multi-qubit controlled $R_{x}$ gates to achieve the same AR as DS-QAOA+ and QLS. Given that NISQ devices are characterized by limited qubit connectivity and high gate noise, the reduction in qubit and multi-qubit controlled gate consumption in PQA significantly enhances its feasibility for near-term quantum hardware.

\medskip
In our work, the heuristic expansion strategies in PQA prioritize nodes that increase the likelihood of retaining the MIS solution while reducing additional quantum resource consumption. To maintain fine-grained control over subgraph growth, PQA starts with a small subgraph and incrementally expands it by adding one node at a time during each expansion step. However, this incremental approach may lead to a higher number of optimization rounds, potentially increasing the overall computational overheads. To address this limitation and improve convergence efficiency, future research should explore more effective graph construction strategies and the optimal choice of initial size, taking into account the graph structure to maintain PQA performance. One possible approach to reducing external optimization rounds is to introduce multiple vertices in each expansion step. However, further research is required to evaluate its impact on the effectiveness and stability of PQA. Another promising direction is to incorporate machine learning techniques to optimize the selection of initial graph size and dynamically refine the graph expansion strategy based on feedback obtained during the optimization process. These advancements could significantly reduce the number of iterations required in the external optimization loop, thereby enhancing the computational efficiency of PQA while improving its scalability to tackle larger problem instances. Finally, although this paper focuses on the MIS problem, the progressive solving framework can also be applied to other combinatorial optimization problems, such as the Maximum Weighted Independent Set problem, the Facility Location Problem, and the Knapsack Problem. A promising direction for future research is to design tailored expansion strategies for different optimization problems and investigate the applicability and performance of PQA in solving these problems.

\medskip
\textbf{Acknowledgements} \par 
Thanks for the Tianyan Quantum Computing Program. This work is supported by National Natural Science Foundation of China (Grant Nos. 62371069, 62372048, 62272056), and BUPT Excellent Ph.D. Students Foundation(CX2023123).

\appendix
\setcounter{figure}{0}
\setcounter{table}{0}
\setcounter{equation}{0}
\renewcommand{\thefigure}{A\arabic{figure}}
\renewcommand{\thetable}{A\arabic{table}}
\renewcommand{\theequation}{A\arabic{equation}}

\section{The extension of the subgraph}\label{graph_extension}
\begin{figure}[htbp]
	\centering
	\setlength{\abovecaptionskip}{0.cm}
	\subfigure{
		\includegraphics[width=0.48\textwidth]{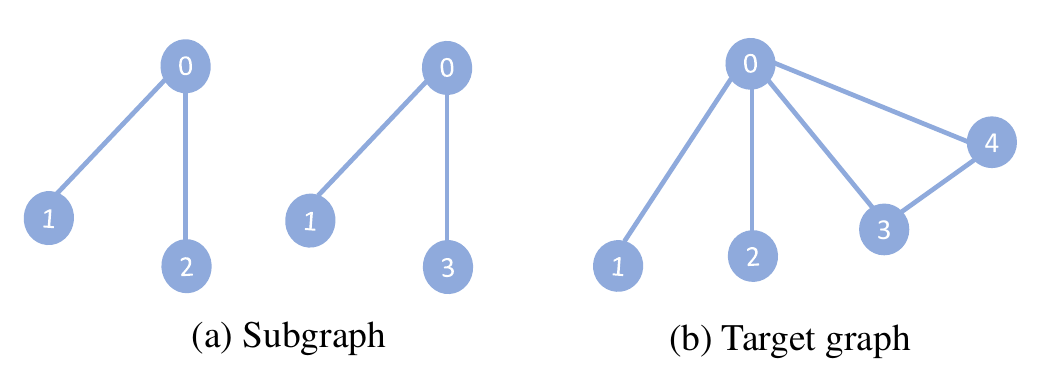}
	}
	\caption{The target graph and its induced subgraphs.}	\label{graph_cases}
\end{figure}

Denote the current subgraph as $G_{i} = (V_{i},E_{i})$. For the first subgraph in FIG.~\ref{graph_cases}, $V_{i} = \left \{ 0,1,2 \right \} $ and $E_{i} = \left \{(0,1),(0,2)  \right \} $. The candidate vertex set  $V_{C} = V-V_{i} = \left \{ 3,4 \right \} $. Our goal is to select the vertex from $V_{C}$ that has the least connection with the current subgraph to be added to $G_{i}$. Specifically, if node 3 is added to $G_{i}$, $C(v_{c} = 3) = 1$. Similarly, if node 4 is added, $C(v_{c} = 4) = 1$. Since both nodes have the same connection, we proceed to the next round of selection. When node 3 is added to $V_{i}$, the edge $(0,3)$ is included in $E_{i}$, leaving $V_{C} = \left \{4 \right \}$ and $C(v_{c} = 4) = 2$. Similarly, if node 4 were added, $C(v_{c} = 3) = 2$. After evaluating both nodes, since they equally satisfy the selection criteria, one can be randomly added to $V_{i}$. If node 3 (or 4) is added, the edge $(0,3)$ (or $(0,4)$) is included in $E_{i}$. The resulting subgraph becomes is $G_{i+1} = \left \{  \left \{ 0,1,2,3 \right \}  ,\left \{(0,1),(0,2),(0,3)  \right \} \right \} $ (or $G_{i+1} = \left \{  \left \{ 0,1,2,4 \right \}  ,\left \{(0,1),(0,2),(0,4)  \right \} \right \}$).

\medskip
For the second subgraph in FIG.~\ref{graph_cases}, $V_{i} = \left \{ 0,1,3 \right \} $ and $E_{i} = \left \{ (0,1),(0,3) \right \} $. The candidate vertex set $V_{C} = \left \{ 2,4 \right \} $, with $C(v_{c} = 2) = 1$ and $C(v_{c} = 4) = 2$. We choose the node with the smallest $C(v_{c})$ value from the candidate vertex set. Thus, node 2 is the next node to be added to $V_{i}$, and the edge $(0,2)$ is added to $E_{i}$. The newly constructed subgraph is $G_{i+1} = \left \{  \left \{ 0,1,2,3 \right \}  ,\left \{(0,1),(0,2),(0,3)  \right \} \right \} $.

\section{The methods of calculating total resources} \label{calculate_resources}
When the algorithm reaches a given AR at a certain level depth, the total resource consumption is calculated as the product of the total number of runs required to achieve the given AR under this level depth and the average resource consumption per run. Therefore, the key to this calculation lies in determining the required level depths, the total number of runs, and the resource consumption per run. To determine at which level depth the algorithm can achieve the given AR, we vary the level depth from 1 to $n$ and perform multiple optimization runs at each depth. Specifically, we conduct 100 optimization runs at each level depth $p$ for $p \le 6$, and $2^{p}$ optimization runs when $p > 6$. After completing all simulations, we first determine the shallowest level depths at which the algorithm achieves the given AR, based on the optimal approximate ratio obtained from multiple optimization runs at each depth. Next, we calculate the total number of runs and average resource consumption at this level depth. The total number of runs required to achieve the given AR is estimated using the following method. First, we calculate the number of runs $Q$ where the resulting AR exceeds the given threshold across multiple optimization runs. Next, we calculate the ratio of $Q$ to the total number of runs at level depth $p$, which represents the probability of exceeding the threshold. The reciprocal of this probability provides the total number of runs needed for the algorithm to achieve the given AR. We then calculate the total resource consumption for each algorithm to achieve the given AR on each graph. Next, we average the total number of runs, iterations, qubits, multi-qubit controlled $R_{x}$ gates, and runtime for each graph type, and use these values to represent the resources required by each algorithm to achieve the given AR on that graph type.

\begin{figure*}[htbp]
	\centering
	\setlength{\abovecaptionskip}{0.cm}
	\subfigure[Total runs, ER with prob = 0.4]{
		\includegraphics[width=0.3\textwidth]{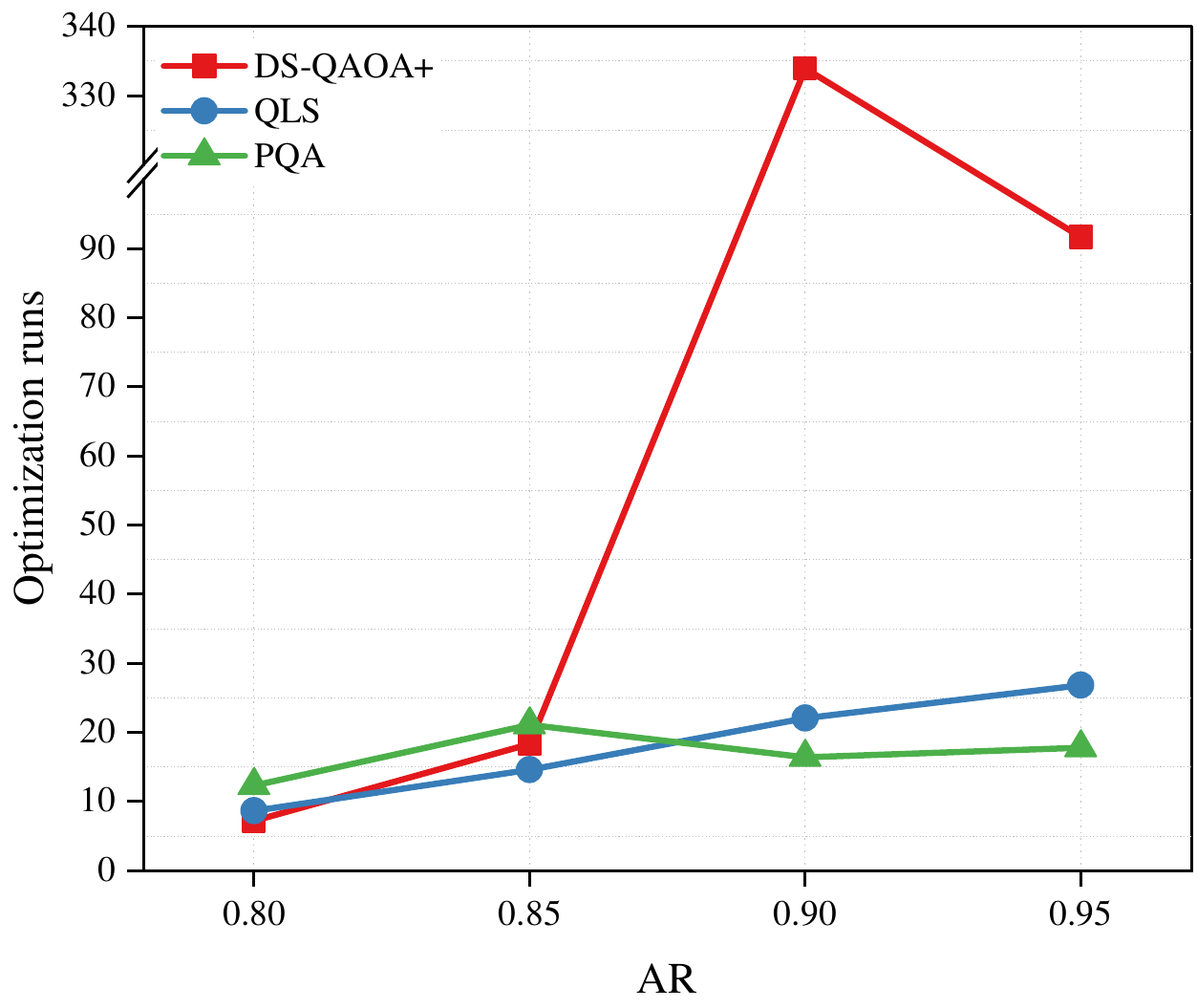}}
	\subfigure[ Total runs, ER with prob = 0.5]{
		\includegraphics[width=0.3\textwidth]{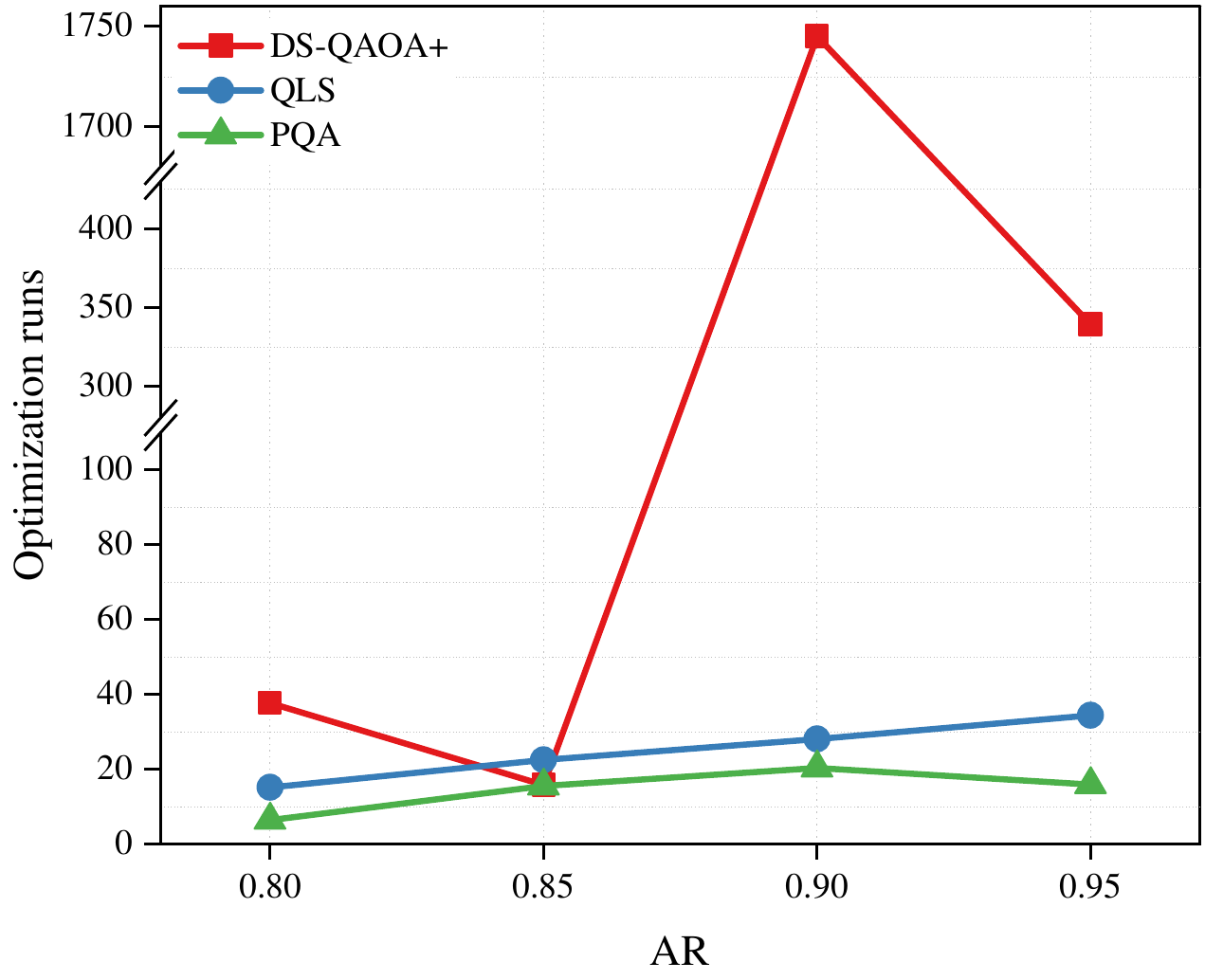}}
        \subfigure[ Total runs, 3-regular graphs]{
		\includegraphics[width=0.3\textwidth]{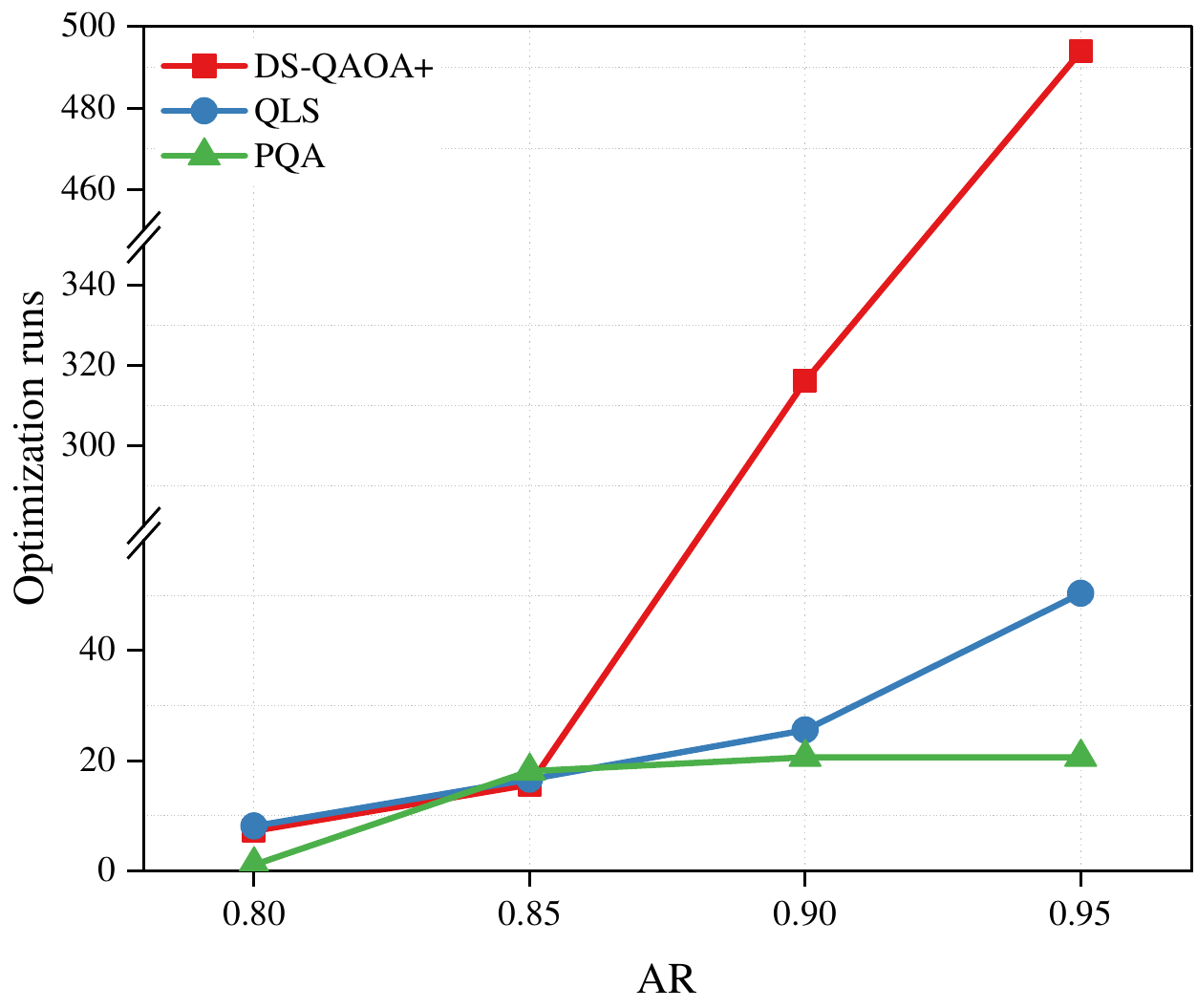}}  
	\caption{The total optimization runs of various algorithms when they achieve the given AR on different graph types.}	\label{runs_consumption}
\end{figure*}

\subsection{The methods of calculating the runtime.} \label{calculate_runtime}
Runtime serves as a comprehensive measure of the execution efficiency of the algorithm, incorporating both quantum and classical computing time. For algorithms like QLS and PQA, classical computations (such as determining edge existence and graph expansion) contribute to the total runtime. In contrast, DS-QAOA+ excludes these computations. Although the time for a single-edge judgment is minimal, we include this classical computation time in our runtime comparison to maintain fairness.

\medskip
For QLS and PQA, the time required for a single optimization run includes both quantum and classical components. The classical time $T_{c}$ is the time consumed in determining whether there is an edge between vertices during updating the global solution or expanding the subgraph. Assuming that the time required for a single determination is $t_{j}$, and the total number of determinations required by the algorithm in an optimization run is $J$, then
\begin{equation}
	T_{c} = J \times t_{j}.\label{T_c}
\end{equation} 

The quantum running time $T_{q}$ includes the time required for initial state preparation, quantum state measurements, and quantum gate evolution across multiple iterations \cite{runtime}. In each iteration, $M$ repeated measurements are needed to calculate the average expectation value of the objective function. The time consumed in a single repetition is 
\begin{equation}
	t_{q} = t_{p} + t_{M} + (\text{circuit depth}) \times t_{G},\label{t_q}
\end{equation} 
where $t_{p}$ is the initial state preparation time, $t_{M}$ is the time for a single measurement, $t_{G}$ is the time for a single quantum gate evolution, and $\text{circuit depth}$ is the depth of the parameterized quantum circuit. The time consumed for a single iteration is 
\begin{equation}
T_{\text{ITR}} = t_{q} \times M,\label{T_ITR}
\end{equation}
and the quantum time consumed in a single optimization run is 
\begin{equation}
	T_{q} = T_{\text{ITR}} \times \text{ITRs},\label{T_q}
\end{equation}
where ITRs are the number of iterations required for the objective function to converge in this run. Therefore, the time consumed in a single run for QLS and PQA is 
\begin{equation}
	T_{\text{PQA}} = T_{\text{q, PQA}} + T_{\text{c, PQA}}, \label{T_PQA}
\end{equation}
\begin{equation}
	T_{\text{QLS}} = T_{\text{q, QLS}} + T_{\text{c, QLS}}, \label{T_QLS}
\end{equation}
respectively, while for DS-QAOA+, its running time is 
\begin{equation}
	T_{\text{DS}} = T_{\text{q, DS}}.\label{T_DS}
\end{equation}

This study explores the runtime required by different algorithms to achieve the same AR across various graph types. The calculations employ the same realistic assumptions as Ref.~\cite{runtime}, namely, $t_{p} + t_{M}$ is one millisecond, and $t_{G}$ is 10 nanoseconds. The tests are conducted on a server equipped with an Intel(R) Xeon(R) Sliver 4316 CPU running at 2.3GHz and 251GB of RAM, where $t_{j}$ is determined to be 0.31226 milliseconds.

\bibliographystyle{elsarticle-num} 
\bibliography{refe}

\begin{thebibliography}{10}
\expandafter\ifx\csname url\endcsname\relax
  \def\url#1{\texttt{#1}}\fi
\expandafter\ifx\csname urlprefix\endcsname\relax\def\urlprefix{URL }\fi
\expandafter\ifx\csname href\endcsname\relax
  \def\href#1#2{#2} \def\path#1{#1}\fi

\bibitem{grover}
L.~K. Grover, Quantum mechanics helps in searching for a needle in a haystack, Phys. Rev. Lett. 79 (1997) 325--328.
\newblock \href {https://doi.org/10.1103/PhysRevLett.79.325} {\path{doi:10.1103/PhysRevLett.79.325}}.

\bibitem{shor}
P.~W. Shor, Polynomial-time algorithms for prime factorization and discrete logarithms on a quantum computer, SIAM Review 41~(2) (1999) 303--332.
\newblock \href {https://doi.org/10.1137/S0036144598347011} {\path{doi:10.1137/S0036144598347011}}.

\bibitem{qpca}
S.~Lloyd, M.~Mohseni, P.~Rebentrost, Quantum principal component analysis, Nature Physics 10~(9) (2014) 631--633.
\newblock \href {https://doi.org/10.1038/nphys3029} {\path{doi:10.1038/nphys3029}}.

\bibitem{reduction_yu}
C.-H. Yu, F.~Gao, Q.-Y. Wen, An improved quantum algorithm for ridge regression, IEEE Transactions on Knowledge and Data Engineering 33~(3) (2019) 858--866.
\newblock \href {https://doi.org/10.1109/TKDE.2019.2937491} {\path{doi:10.1109/TKDE.2019.2937491}}.

\bibitem{reduction_pan}
S.-J. Pan, L.-C. Wan, H.-L. Liu, Q.-L. Wang, S.-J. Qin, Q.-Y. Wen, F.~Gao, Improved quantum algorithm for a-optimal projection, Phys. Rev. A 102 (2020) 052402.
\newblock \href {https://doi.org/10.1103/PhysRevA.102.052402} {\path{doi:10.1103/PhysRevA.102.052402}}.

\bibitem{reduction_wan}
L.-C. Wan, C.-H. Yu, S.-J. Pan, F.~Gao, Q.-Y. Wen, S.-J. Qin, Asymptotic quantum algorithm for the toeplitz systems, Phys. Rev. A 97 (2018) 062322.
\newblock \href {https://doi.org/10.1103/PhysRevA.97.062322} {\path{doi:10.1103/PhysRevA.97.062322}}.

\bibitem{Lijing}
J.~Li, F.~Gao, S.~Lin, M.~Guo, Y.~Li, H.~Liu, S.~Qin, Q.~Wen, Quantum k-fold cross-validation for nearest neighbor classification algorithm, Physica A: Statistical Mechanics and its Applications 611 (2023) 128435.
\newblock \href {https://doi.org/10.1016/j.physa.2022.128435} {\path{doi:10.1016/j.physa.2022.128435}}.

\bibitem{hhl}
A.~W. Harrow, A.~Hassidim, S.~Lloyd, Quantum algorithm for linear systems of equations, Phys. Rev. Lett. 103 (2009) 150502.
\newblock \href {https://doi.org/10.1103/PhysRevLett.103.150502} {\path{doi:10.1103/PhysRevLett.103.150502}}.

\bibitem{NISQ}
J.~Preskill, Quantum computing in the {NISQ} era and beyond, Quantum 2 (2018) 79.
\newblock \href {https://doi.org/10.22331/q-2018-08-06-79} {\path{doi:10.22331/q-2018-08-06-79}}.

\bibitem{VQA_review1}
M.~Cerezo, A.~Arrasmith, R.~Babbush, S.~C. Benjamin, S.~Endo, K.~Fujii, J.~R. McClean, K.~Mitarai, X.~Yuan, L.~Cincio, P.~J. Coles, Variational quantum algorithms, Nature Reviews Physics 3~(9) (2021) 625--644.
\newblock \href {https://doi.org/10.1038/s42254-021-00348-9} {\path{doi:10.1038/s42254-021-00348-9}}.

\bibitem{VQA_review2}
K.~Bharti, A.~Cervera-Lierta, T.~H. Kyaw, T.~Haug, S.~Alperin-Lea, A.~Anand, M.~Degroote, H.~Heimonen, J.~S. Kottmann, T.~Menke, et~al., Noisy intermediate-scale quantum algorithms, Reviews of Modern Physics 94~(1) (2022) 015004.
\newblock \href {https://doi.org/10.1103/RevModPhys.94.015004} {\path{doi:10.1103/RevModPhys.94.015004}}.

\bibitem{adaptive-vqe}
H.~R. Grimsley, S.~E. Economou, E.~Barnes, N.~J. Mayhall, An adaptive variational algorithm for exact molecular simulations on a quantum computer, Nature communications 10~(1) (2019) 3007.
\newblock \href {https://doi.org/10.1038/s41467-019-10988-2} {\path{doi:10.1038/s41467-019-10988-2}}.

\bibitem{qubit-adaptive-vqe}
H.~L. Tang, V.~Shkolnikov, G.~S. Barron, H.~R. Grimsley, N.~J. Mayhall, E.~Barnes, S.~E. Economou, qubit-adapt-vqe: An adaptive algorithm for constructing hardware-efficient ans{\"a}tze on a quantum processor, PRX Quantum 2~(2) (2021) 020310.
\newblock \href {https://doi.org/10.1103/PRXQuantum.2.020310} {\path{doi:10.1103/PRXQuantum.2.020310}}.

\bibitem{QCL}
D.~Zhu, N.~M. Linke, M.~Benedetti, K.~A. Landsman, N.~H. Nguyen, C.~H. Alderete, A.~Perdomo-Ortiz, N.~Korda, A.~Garfoot, C.~Brecque, et~al., Training of quantum circuits on a hybrid quantum computer, Science advances 5~(10) (2019) eaaw9918.
\newblock \href {https://doi.org/10.1126/sciadv.aaw9918} {\path{doi:10.1126/sciadv.aaw9918}}.

\bibitem{QAS}
Y.~Du, T.~Huang, S.~You, M.-H. Hsieh, D.~Tao, Quantum circuit architecture search for variational quantum algorithms, npj Quantum Information 8~(1) (2022) 62.
\newblock \href {https://doi.org/10.1038/s41534-022-00570-y} {\path{doi:10.1038/s41534-022-00570-y}}.

\bibitem{RE-ADAPT-VQA}
S.~Wu, Y.~Song, R.~Li, S.-J. Qin, Q.~Wen, F.~Gao, Resource‐efficient adaptive variational quantum algorithm for combinatorial optimization problems, Advanced Quantum Technologies (01 2025).
\newblock \href {https://doi.org/10.1002/qute.202400484} {\path{doi:10.1002/qute.202400484}}.

\bibitem{QFL}
Y.~Song, Y.~Wu, S.~Wu, D.~Li, Q.~Wen, S.~Qin, F.~Gao, A quantum federated learning framework for classical clients, Science China Physics, Mechanics \& Astronomy 67~(5) (2024) 250311.
\newblock \href {https://doi.org/10.1007/s11433-023-2337-2} {\path{doi:10.1007/s11433-023-2337-2}}.

\bibitem{wu_quantum}
Y.~Wu, B.~Wu, J.~Wang, X.~Yuan, Quantum phase recognition via quantum kernel methods, Quantum 7 (2023) 981.
\newblock \href {https://doi.org/10.22331/q-2023-04-17-981} {\path{doi:10.22331/q-2023-04-17-981}}.

\bibitem{wu_qst}
Y.~Wu, Z.~Huang, J.~Sun, X.~Yuan, J.~B. Wang, D.~Lv, Orbital expansion variational quantum eigensolver, Quantum Science and Technology (2023).
\newblock \href {https://doi.org/10.1088/2058-9565/acf9c7} {\path{doi:10.1088/2058-9565/acf9c7}}.

\bibitem{qaoa}
E.~Farhi, J.~Goldstone, S.~Gutmann, A quantum approximate optimization algorithm (2014).
\newblock \href {http://arxiv.org/abs/1411.4028} {\path{arXiv:1411.4028}}, \href {https://doi.org/10.48550/arXiv.1411.4028} {\path{doi:10.48550/arXiv.1411.4028}}.

\bibitem{RQAOA}
S.~Bravyi, A.~Kliesch, R.~Koenig, E.~Tang, Obstacles to variational quantum optimization from symmetry protection, Physical review letters 125~(26) (2020) 260505.
\newblock \href {https://doi.org/10.1103/PhysRevLett.125.260505} {\path{doi:10.1103/PhysRevLett.125.260505}}.

\bibitem{tree-QAOA}
M.~Streif, M.~Leib, Training the quantum approximate optimization algorithm without access to a quantum processing unit, Quantum Science and Technology 5~(3) (2020) 034008.
\newblock \href {https://doi.org/10.1088/2058-9565/ab8c2b} {\path{doi:10.1088/2058-9565/ab8c2b}}.

\bibitem{adaptive-qaoa}
L.~Zhu, H.~L. Tang, G.~S. Barron, F.~A. Calderon-Vargas, N.~J. Mayhall, E.~Barnes, S.~E. Economou, Adaptive quantum approximate optimization algorithm for solving combinatorial problems on a quantum computer, Phys. Rev. Res. 4 (2022) 033029.
\newblock \href {https://doi.org/10.1103/PhysRevResearch.4.033029} {\path{doi:10.1103/PhysRevResearch.4.033029}}.

\bibitem{xiumei}
X.~Zhao, Y.~Li, J.~Li, S.~Wang, S.~Wang, S.~Qin, F.~Gao, Near-term quantum algorithm for solving the maxcut problem with fewer quantum resources, Physica A: Statistical Mechanics and its Applications 648 (2024) 129951.
\newblock \href {https://doi.org/10.1016/j.physa.2024.129951} {\path{doi:10.1016/j.physa.2024.129951}}.

\bibitem{exact_cover}
P.~Vikst\aa{}l, M.~Gr\"onkvist, M.~Svensson, M.~Andersson, G.~Johansson, G.~Ferrini, Applying the quantum approximate optimization algorithm to the tail-assignment problem, Phys. Rev. Appl. 14 (2020) 034009.
\newblock \href {https://doi.org/10.1103/PhysRevApplied.14.034009} {\path{doi:10.1103/PhysRevApplied.14.034009}}.

\bibitem{INTERP}
L.~Zhou, S.-T. Wang, S.~Choi, H.~Pichler, M.~D. Lukin, Quantum approximate optimization algorithm: Performance, mechanism, and implementation on near-term devices, Phys. Rev. X 10 (2020) 021067.
\newblock \href {https://doi.org/10.1103/PhysRevX.10.021067} {\path{doi:10.1103/PhysRevX.10.021067}}.

\bibitem{perform1}
J.~Wurtz, P.~Love, Maxcut quantum approximate optimization algorithm performance guarantees for $p\&gt;1$, Phys. Rev. A 103 (2021) 042612.
\newblock \href {https://doi.org/10.1103/PhysRevA.103.042612} {\path{doi:10.1103/PhysRevA.103.042612}}.

\bibitem{constraints_QAOA}
Y.~Ruan, Z.~Yuan, X.~Xue, Z.~Liu, Quantum approximate optimization for combinatorial problems with constraints, Information Sciences 619 (2023) 98--125.
\newblock \href {https://doi.org/10.1016/j.ins.2022.11.020} {\path{doi:10.1016/j.ins.2022.11.020}}.

\bibitem{portfolio}
S.~Brandhofer, D.~Braun, V.~Dehn, G.~Hellstern, M.~H{\"u}ls, Y.~Ji, I.~Polian, A.~S. Bhatia, T.~Wellens, Benchmarking the performance of portfolio optimization with qaoa, Quantum Information Processing 22~(1) (2022) 25.
\newblock \href {https://doi.org/10.1007/s11128-022-03766-5} {\path{doi:10.1007/s11128-022-03766-5}}.

\bibitem{divide_MIS}
T.~Tomesh, Z.~H. Saleem, M.~A. Perlin, P.~Gokhale, M.~Suchara, M.~Martonosi, Divide and conquer for combinatorial optimization and distributed quantum computation, in: 2023 IEEE International Conference on Quantum Computing and Engineering (QCE), Vol.~1, IEEE, 2023, pp. 1--12.
\newblock \href {https://doi.org/10.1109/QCE57702.2023.00009} {\path{doi:10.1109/QCE57702.2023.00009}}.

\bibitem{MLI}
X.-H. Ni, B.-B. Cai, H.-L. Liu, S.-J. Qin, F.~Gao, Q.-Y. Wen, Multilevel leapfrogging initialization strategy for quantum approximate optimization algorithm, Advanced Quantum Technologies 7~(5) (2024) 2300419.
\newblock \href {https://doi.org/10.1002/qute.202300419} {\path{doi:10.1002/qute.202300419}}.

\bibitem{minimum_vertex_cover}
Y.~Zhang, X.~Mu, X.~Liu, X.~Wang, X.~Zhang, K.~Li, T.~Wu, D.~Zhao, C.~Dong, Applying the quantum approximate optimization algorithm to the minimum vertex cover problem, Applied Soft Computing 118 (2022) 108554.
\newblock \href {https://doi.org/10.1016/j.asoc.2022.108554} {\path{doi:10.1016/j.asoc.2022.108554}}.

\bibitem{QIRO}
J.~R. Fin\ifmmode~\check{z}\else \v{z}\fi{}gar, A.~Kerschbaumer, M.~J. Schuetz, C.~B. Mendl, H.~G. Katzgraber, Quantum-informed recursive optimization algorithms, PRX Quantum 5 (2024) 020327.
\newblock \href {https://doi.org/10.1103/PRXQuantum.5.020327} {\path{doi:10.1103/PRXQuantum.5.020327}}.

\bibitem{Ising_formulations}
A.~Lucas, Ising formulations of many np problems, Frontiers in physics 2 (2014) 5.

\bibitem{RUAN}
Y.~Ruan, Z.~Yuan, X.~Xue, Z.~Liu, Quantum approximate optimization for combinatorial problems with constraints, Information Sciences 619 (2023) 98--125.
\newblock \href {https://doi.org/10.1016/j.ins.2022.11.020} {\path{doi:10.1016/j.ins.2022.11.020}}.

\bibitem{alternating_QAOA_ansatz}
S.~Hadfield, Z.~Wang, B.~O’gorman, E.~G. Rieffel, D.~Venturelli, R.~Biswas, From the quantum approximate optimization algorithm to a quantum alternating operator ansatz, Algorithms 12~(2) (2019) 34.
\newblock \href {https://doi.org/10.3390/a12020034} {\path{doi:10.3390/a12020034}}.

\bibitem{graph_coloring1}
S.~Hadfield, Z.~Wang, E.~G. Rieffel, B.~O'Gorman, D.~Venturelli, R.~Biswas, Quantum approximate optimization with hard and soft constraints, in: Proceedings of the Second International Workshop on Post Moores Era Supercomputing, PMES'17, Association for Computing Machinery, New York, NY, USA, 2017, p. 15–21.
\newblock \href {https://doi.org/10.1145/3149526.3149530} {\path{doi:10.1145/3149526.3149530}}.

\bibitem{graph_coloring2}
Z.~Wang, N.~C. Rubin, J.~M. Dominy, E.~G. Rieffel, X y mixers: Analytical and numerical results for the quantum alternating operator ansatz, Physical Review A 101~(1) (2020) 012320.
\newblock \href {https://doi.org/10.1103/PhysRevA.101.012320} {\path{doi:10.1103/PhysRevA.101.012320}}.

\bibitem{portfolio_optimization}
Z.~He, R.~Shaydulin, S.~Chakrabarti, D.~Herman, C.~Li, Y.~Sun, M.~Pistoia, Alignment between initial state and mixer improves qaoa performance for constrained optimization, npj Quantum Information 9~(1) (2023) 121.
\newblock \href {https://doi.org/10.1038/s41534-023-00787-5} {\path{doi:10.1038/s41534-023-00787-5}}.

\bibitem{MIS}
Z.~H. Saleem, Max-independent set and the quantum alternating operator ansatz, International Journal of Quantum Information 18~(04) (2020) 2050011.
\newblock \href {https://doi.org/10.1142/S0219749920500112} {\path{doi:10.1142/S0219749920500112}}.

\bibitem{minimum_exact_cover}
S.-S. Wang, H.-L. Liu, Y.-Q. Song, F.~Gao, S.-J. Qin, Q.-Y. Wen, Quantum alternating operator ansatz for solving the minimum exact cover problem, Physica A: Statistical Mechanics and its Applications 626 (2023) 129089.
\newblock \href {https://doi.org/10.1016/j.physa.2023.129089} {\path{doi:10.1016/j.physa.2023.129089}}.

\bibitem{QLS}
T.~Tomesh, Z.~H. Saleem, M.~Suchara, Quantum local search with the quantum alternating operator ansatz, Quantum 6 (2022) 781.
\newblock \href {https://doi.org/10.22331/q-2022-08-22-781} {\path{doi:10.22331/q-2022-08-22-781}}.

\bibitem{MIS_in_network}
T.~Moscibroda, R.~Wattenhofer, Maximal independent sets in radio networks, in: Proceedings of the Twenty-Fourth Annual ACM Symposium on Principles of Distributed Computing, PODC '05, Association for Computing Machinery, New York, NY, USA, 2005, p. 148–157.
\newblock \href {https://doi.org/10.1145/1073814.1073842} {\path{doi:10.1145/1073814.1073842}}.

\bibitem{MIS_in_scheduling}
D.~Eddy, M.~J. Kochenderfer, A maximum independent set method for scheduling earth-observing satellite constellations, Journal of Spacecraft and Rockets 58~(5) (2021) 1416--1429.
\newblock \href {https://doi.org/10.2514/1.A34931} {\path{doi:10.2514/1.A34931}}.

\bibitem{MIS_vertex_cover}
D.~S. Hochbaum, Approximating covering and packing problems: set cover, vertex cover, independent set, and related problems, Approximation algorithms for NP-hard problems (1997) 94--143.

\bibitem{PIL}
L.~Li, J.~Li, Y.~Song, S.~Qin, Q.~Wen, F.~Gao, An efficient quantum proactive incremental learning algorithm, Science China Physics, Mechanics \& Astronomy (2024).
\newblock \href {https://doi.org/10.1007/s11433-024-2501-4} {\path{doi:10.1007/s11433-024-2501-4}}.

\bibitem{reducibility}
R.~M. Karp, Reducibility among combinatorial problems, Springer, 2010.

\bibitem{Iterative_MIS}
L.~T. Brady, S.~Hadfield, Iterative quantum algorithms for maximum independent set, Phys. Rev. A 110 (2024) 052435.
\newblock \href {https://doi.org/10.1103/PhysRevA.110.052435} {\path{doi:10.1103/PhysRevA.110.052435}}.

\bibitem{backtracking_algorithm_MIS}
E.~Loukakis, A new backtracking algorithm for generating the family of maximal independent sets of a graph, Computers \& Mathematics with Applications 9~(4) (1983) 583--589.
\newblock \href {https://doi.org/10.1016/0898-1221(83)90115-3} {\path{doi:10.1016/0898-1221(83)90115-3}}.

\bibitem{branch_and_bound_MIS}
J.~S. Warren, I.~V. Hicks, Combinatorial branch-and-bound for the maximum weight independent set problem, Relat{\'o}rio T{\'e}cnico, Texas A\&M University, Citeseer 9 (2006) 17.

\bibitem{exact_algorithm_MIS}
M.~Xiao, H.~Nagamochi, Exact algorithms for maximum independent set, Information and Computation 255 (2017) 126--146.
\newblock \href {https://doi.org/10.1016/j.ic.2017.06.001} {\path{doi:10.1016/j.ic.2017.06.001}}.

\bibitem{evolutionary_algorithm_MIS}
T.~Back, S.~Khuri, An evolutionary heuristic for the maximum independent set problem, in: Proceedings of the first IEEE conference on evolutionary computation. IEEE World Congress on Computational Intelligence, IEEE, 1994, pp. 531--535.
\newblock \href {https://doi.org/10.1109/ICEC.1994.350004} {\path{doi:10.1109/ICEC.1994.350004}}.

\bibitem{tabu_search}
R.~Aringhieri, et~al., A tabu search algorithm for solving chance-constrained programs, Journal of the ACM 5 (2004) 1--14.

\bibitem{hybrid_iterated_LS}
B.~Nogueira, R.~G. Pinheiro, A.~Subramanian, A hybrid iterated local search heuristic for the maximum weight independent set problem, Optimization Letters 12 (2018) 567--583.
\newblock \href {https://doi.org/10.1007/s11590-017-1128-7} {\path{doi:10.1007/s11590-017-1128-7}}.

\bibitem{iterated_LS}
B.~Nogueira, R.~Pinheiro, E.~Tavares, Iterated local search for the generalized independent set problem, Optimization Letters 15 (06 2021).
\newblock \href {https://doi.org/10.1007/s11590-020-01643-7} {\path{doi:10.1007/s11590-020-01643-7}}.

\bibitem{local_search_MIS}
D.~V. Andrade, M.~G. Resende, R.~F. Werneck, Fast local search for the maximum independent set problem, Journal of Heuristics 18 (2012) 525--547.
\newblock \href {https://doi.org/10.1007/s10732-012-9196-4} {\path{doi:10.1007/s10732-012-9196-4}}.

\bibitem{gate_decomposition}
R.~Vale, T.~M.~D. Azevedo, I.~Ara{\'u}jo, I.~F. Araujo, A.~J. da~Silva, Decomposition of multi-controlled special unitary single-qubit gates, arXiv preprint arXiv:2302.06377 (2023).
\newblock \href {https://doi.org/10.48550/arXiv.2302.06377} {\path{doi:10.48550/arXiv.2302.06377}}.

\bibitem{gate_decomposition_MIS}
T.~Tomesh, N.~Allen, D.~Dilley, Z.~Saleem, Quantum-classical tradeoffs and multi-controlled quantum gate decompositions in variational algorithms, {Quantum} 8 (2024) 1493.
\newblock \href {https://doi.org/10.22331/q-2024-10-04-1493} {\path{doi:10.22331/q-2024-10-04-1493}}.

\bibitem{parameter_transferability}
I.~Lyngfelt, L.~Garc\'{\i}a-\'Alvarez, Symmetry-informed transferability of optimal parameters in the quantum approximate optimization algorithm, Phys. Rev. A 111 (2025) 022418.
\newblock \href {https://doi.org/10.1103/PhysRevA.111.022418} {\path{doi:10.1103/PhysRevA.111.022418}}.

\bibitem{parameter_transfer}
R.~Shaydulin, P.~C. Lotshaw, J.~Larson, J.~Ostrowski, T.~S. Humble, \href{https://doi.org/10.1145/3584706}{Parameter transfer for quantum approximate optimization of weighted maxcut}, ACM Transactions on Quantum Computing 4~(3) (Apr. 2023).
\newblock \href {https://doi.org/10.1145/3584706} {\path{doi:10.1145/3584706}}.
\newline\urlprefix\url{https://doi.org/10.1145/3584706}

\bibitem{transfer_learning2024}
J.~A. Montanez-Barrera, D.~Willsch, K.~Michielsen, Transfer learning of optimal qaoa parameters in combinatorial optimization (2024).
\newblock \href {http://arxiv.org/abs/2402.05549} {\path{arXiv:2402.05549}}, \href {https://doi.org/10.48550/arXiv.2402.05549} {\path{doi:10.48550/arXiv.2402.05549}}.

\bibitem{angle_conjecture}
J.~Wurtz, D.~Lykov, Fixed-angle conjectures for the quantum approximate optimization algorithm on regular maxcut graphs, Phys. Rev. A 104 (2021) 052419.
\newblock \href {https://doi.org/10.1103/PhysRevA.104.052419} {\path{doi:10.1103/PhysRevA.104.052419}}.

\bibitem{leapfrogging}
F.~G. S.~L. Brandao, M.~Broughton, E.~Farhi, S.~Gutmann, H.~Neven, For fixed control parameters the quantum approximate optimization algorithm's objective function value concentrates for typical instances (2018).
\newblock \href {http://arxiv.org/abs/1812.04170} {\path{arXiv:1812.04170}}, \href {https://doi.org/10.48550/arXiv.1812.04170} {\path{doi:10.48550/arXiv.1812.04170}}.

\bibitem{hill_climbing}
S.~Chinnasamy, M.~Ramachandran, M.~Amudha, K.~Ramu, A review on hill climbing optimization methodology, Recent Trends in Management and Commerce 3~(1) (2022) 1--7.
\newblock \href {https://doi.org/10.46632/rmc/3/1/1} {\path{doi:10.46632/rmc/3/1/1}}.

\bibitem{runtime}
G.~G. Guerreschi, A.~Y. Matsuura, Qaoa for max-cut requires hundreds of qubits for quantum speed-up, Scientific reports 9~(1) (2019) 6903.
\newblock \href {https://doi.org/10.1038/s41598-019-43176-9} {\path{doi:10.1038/s41598-019-43176-9}}.

\end{thebibliography}
\end{document}